\documentclass{article}

\usepackage{arxiv}

\usepackage[utf8]{inputenc} 
\usepackage[T1]{fontenc}    
\usepackage{hyperref}       
\usepackage{url}            
\usepackage{booktabs}       
\usepackage{amsfonts}       
\usepackage{nicefrac}       
\usepackage{microtype}      
\usepackage{lipsum}

\usepackage[utf8]{inputenc}
\usepackage[T1]{fontenc}
\usepackage{lmodern}
\usepackage{cite}
\usepackage{amsmath,amssymb,amsfonts}
\usepackage{algorithmic}
\usepackage{verbatim}
\usepackage{adjustbox}
\usepackage[normalem]{ulem}

\usepackage{soul, color}
\usepackage{balance}
\usepackage{graphicx}
\usepackage{subfigure}
\graphicspath{ {images/} }
\usepackage{caption,setspace}
\usepackage{textcomp}
\def\BibTeX{{\rm B\kern-.05em{\sc i\kern-.025em b}\kern-.08em
    T\kern-.1667em\lower.7ex\hbox{E}\kern-.125emX}}

\usepackage[nolist]{acronym}

\title{Integrated Methodology to Cognitive Network \& Slice Management in Virtualized  5G Networks}

\author{
  Xenofon Vasilakos \\
  Department of Electrical \& Electronic Engineering, \\
    University of Bristol, Clifton BS8 1UB, UK\\
  \texttt{xenofon.vasilakos@bristol.ac.uk}, \\
  Mobile Communications Department, \\
  EURECOM, 06410, Biot, France\\
  \texttt{xenofon.vasilakos@eurecom.fr} 
\And
  Navid Nikaein \\
  Mobile Communications Department, \\
  EURECOM, 06410, Biot, France\\
  \texttt{navid.nikaein@eurecom.fr} \\
\And  
 Dean H Lorenz \\
  IBM Haifa Research Labs,\\ 
  Haifa, Israel \\
  \texttt{dean@il.ibm.com} \\
\And
  Berkay K\"oksal \\
  Mobile Communications Department, \\
  EURECOM, 06410, Biot, France\\
  \texttt{berkay.koksal@eurecom.fr} \\
\And
  Nasim Ferdosian \\
  Mobile Communications Department, \\
  EURECOM, 06410, Biot, France\\
  \texttt{nasim.ferdosian@eurecom.fr} \\
}

\newcommand{\wrt}{w.r.t. }

\newcommand{\ml}{\ac{ml} }

\newcommand{\qos}{\ac{qos} }
\newcommand{\qoe}{\ac{qoe} }
\newcommand{\ete}{\ac{e2e} }
\newcommand{\aka}{a.k.a. }
\newcommand{\ucase}{use case }

\newcommand{\eH}{eHealth }
\newcommand{\pcs}{\ac{pcs} }
\newcommand{\fcsm}{\ac{fcsm} }

\newcounter{Qcountnum}

\newcommand\nq[3]{
	{
      \stepcounter{Qcount\roman{Qcountnum}}
      \textbf{\color{red}\#Quest.\_\csname theQcount\roman{Qcountnum}\endcsname } \color{red}$<$From:#1 To:#2$>$: {\color{red}\hl{#3}}
    }
}

\newcounter{Rcountnum}

\newcommand\nr[1]{
	{\color{red}
      \stepcounter{Rcount\roman{Rcountnum}}
      \#note\_\csname theRcount\roman{Rcountnum}\endcsname:[[#1]]}}

\newcounter{Bcountnum}

\newcommand\nb[1]{
	{\color{blue}
      \stepcounter{Bcount\roman{Bcountnum}}
      \#note\_\csname theBcount\roman{Bcountnum}\endcsname:[[#1]]}}

\begin{document}
\begin{acronym}

\acro{zsm}[ZSM] {Zero touch network \& Service Management}

\acro{pcs}[PCS]{Proactive Control Scheme}
\acro{fcsm}[4-CSM]{Four-stage approach to Cognitive Slice Management}

\acro{poc}[PoC]{Proof-of-Concept}

\acro{iot}[IoT]{Internet-of-Things}

\acro{oai}[OAI]{OpenAirInterface}

\acro{cqi}[CQI]{Channel Quality Indicator}
\acro{wbcqi}[wbCQI]{wideband CQI}
\acro{sinr}[SINR]{signal-to-interference plus noise ratio}
\acro{rb}[RB]{Resource Block}

\acro{capex}[CAPEX]{Capital Expenditure}
\acro{opex}[OPEX]{Operating Expenses}

\acro{no}[NO]{Network Operator}
\acro{vno}[VNO]{Virtual Network Operators}
\acro{csp}[CSP]{Communication Service Provider}

\acro{vnf}[VNF]{Virtual Network Function}
\acro{vna}[VNA]{Virtual Network Applications}
\acro{sdn}[SDN]{Software-Defined Networking}
\acro{mec}[MEC]{Multi-access Edge Computing}
\acro{llmec}[LL-MEC]{Low Latency-MEC}

\acro{cpu}[CPU]{Central Processor Unit}
\acro{gpu}[GPU]{Graphical Processing Unit}
\acro{tpu}[TPU]{Tensor Processing Unit}

\acro{gnb}[gNB]{next Generation NodeB}

\acro{sla}[SLA]{Service Level Agreement}

\acro{ue}[UE]{User Equipment}

\acro{wrt}[w.r.t.]{with respect to}
\acro{rss}[RSS]{Received Signal Strength}

\acro{qos}[QoS]{Quality-of-Service}
\acro{qoe}[QoE]{Quality-of-Experience}
\acro{e2e}[E2E]{End-to-End}

\acro{kb}[KB]{Knowledge Base}

\acro{bh}[BH]{BackHaul}
\acro{fh}[FH]{FrontHaul}

\acro{ran}[RAN]{Radio Access Network}
\acro{sdran}[SD-RAN]{Software-Defined Radio Access Network }

\acro{cn}[CN]{Core Network}

\acro{ml}[ML]{Machine Learning}

\acro{lstm}[LSTM]{Long-Short Term Memory}
\acro{ann}[ANN]{Artificial Neural Network}

\acro{epc}[EPC]{Evolved Packet Core}
\acro{gtp}[GTP]{GPRS Tunneling Protocol}

\acro{wrt}[wrt]{with respect to}
\acro{etxi}[ETSI]{European Telecommunications Standard Institute}

\acro{psn}[PSN]{Public Safety Networks }
\acro{upf}[UPF]{User-Plane Function}
\acro{cnsm}[CNSM]{Cognitive Network \& Slice Management}

\acro{lasso}[LASSO]{Least Absolute Shrinkage and Selection Operator}
\acro{xgboost}[XGBoost]{Extreme Gradient Boosting}

\acro{mac}[MAC]{Medium Access Control}
\acro{rrc}[RRC]{Radio Resource Control}
\acro{pdcp}[PDCP]{Packet Data Convergence Protocol}

\acro{rsrq}[RSRQ]{Reference Signal Received Quality}
\acro{rsrp}[RSRP]{Reference Signal Received Power}
\acro{phr}[PHR]{Power Headroom Report}
\acro{urllc}[URLLC]{Ultra-Reliable Low-Latency Communication}

\acro{rmse}[RMSE]{Root Mean Square Error}
\acro{mape}[MAPE]{Mean Absolute Percentage Error}

\acro{MAPEk}[MAPE-K]{Monitor, Analyze, Plan and Execute over
a shared Knowledge}

\acro{mie}[MIE]{Massive Emergency Event}

\acro{hd}[HD]{High Definition}
\acro{sd}[SD]{Standard Definition}

\acro{ppdr}[PPDR]{Public Protection and Disaster Relief}

\acro{son}[SON]{Self-Organizing Networks}

\end{acronym}

\maketitle
\begin{abstract}
Fifth Generation (5G) networks are envisioned to be fully autonomous in accordance to the ETSI-defined Zero touch network and Service Management (ZSM) concept. To this end, purpose-specific Machine Learning (ML) models can be used to manage and control physical as well as virtual network resources in a way that is fully compliant to slice Service Level Agreements (SLAs), while also boosting the revenue of the underlying physical network operator(s). This is because specially designed and trained ML models can be both proactive and very effective against slice management issues that can induce significant SLA penalties or runtime costs. However, reaching that point is very challenging. 5G networks will be highly dynamic and complex, offering a large scale of heterogeneous, sophisticated and resource-demanding 5G services as network slices. 
This raises a need for a well-defined, generic and step-wise roadmap to designing, building and deploying efficient ML models as collaborative components of what can be defined as Cognitive Network and Slice Management (CNSM) 5G systems. To address this need, we take a use case-driven approach to design and present a novel Integrated Methodology for CNSM in virtualized 5G networks based on a concrete eHealth use case, and elaborate on it to derive a generic approach for 5G slice management use cases. The three fundamental components that comprise our proposed methodology include (i) a 5G Cognitive Workflow model that conditions everything from the design up to the final deployment of ML models; (ii) a Four-stage approach to Cognitive Slice Management with an emphasis on anomaly detection; and (iii) a Proactive Control Scheme for the collaboration of different ML models targeting different slice life-cycle management problems. 
\end{abstract}

\keywords{5G \and Mobile communication \and Computer network management and control \and Computer networks \and Machine Learning}

\section{Introduction}
\label{sec:introduction}
FIfth Generation (5G) mobile networks pose a major paradigm shift, aimed to improve efficiency and flexibility with a service-oriented architecture that delivers \textit{networks as-a-service}. The underlying concept is to support multiple services and virtual networks
over one or more physical network infrastructure, \ac{wrt} different service definitions and performance requirements. This service-oriented 5G vision can address the vast variety of emerging resource-hungry wireless services~\cite{cisco2017}, via a \textit{cost-efficient} network composition and resource sharing model 
that reduces both \ac{capex} and \ac{opex}. The later is done by decoupling infrastructure providers (e.g., operators and data center owners), 
service providers (e.g., operators and verticals) and network function providers (e.g., vendors).
Therefore, a 5G service can be built by combining multi-vendor physical network functions and \acfp{vnf}, bringing network slicing to the foreground as a key enabler for the envisioned service-oriented 5G \cite{zhang2017network}. 

Slicing enables the composition of multiple logical networks and their delivery-as-a-service or as-a-slice over a shared physical infrastructure \cite{li20175g}. 
A slice can either be completely isolated from the other slices down to the different sets of spectrum and cellular site, or be \textit{shared} across all types of resources including radio spectrum and network functions, or be customized for a subset of user plane and control plane processing with an access to a portion of radio resources in a \textit{virtualized} form. 
Furthermore, a slice may span across multiple domain-specific resources each with different levels of isolation and sharing to accommodate the needs of both \acp{no} and slice owners.

But apart from its grand potential, slicing adds further to the already increased complexity of network and traffic management in 5G.  
The denser and more heterogeneous~\cite{imranZ14, luong18} nature of 5G networks, on the one hand, and the required slice resource provisioning and performance fulfillment, on the other, both call for \acf{zsm}\footnote{
    \ac{zsm} is a concept defined by the European Telecommunications Standards Institute: {https://www.etsi.org/technologies/zero-touch-network-service-management}.
}.
Following a long period of remaining in obscurity, \ac{ml}-based approaches have created a trend towards different aspects of \ac{zsm} in the literature such as discussed in~\cite{wang2017, imranZ14} thanks to breakthroughs made on computational devices, mainly multi-core \acp{cpu}, \acp{gpu} and \acp{tpu}\footnote{
    Tensor Processing Unit by Google: 
    https://goo.gl/8TZLRS
}. 
Therefore \ac{ml} models qualify as an appropriate option for managing multiple coexisting 5G slices on top of physical resources within a unified \acf{cnsm} system.

\subsection{Motivation for an Integrated Methodology}
\label{sec:intro-motiv}

\ml algorithms can be trained to learn how to efficiently tackle \textit{concrete} problems towards (re)organizing network and slice resources \textit{straight} after online input data. This means that even after their deployment the models can keep learning from feedback such as network statistics. Also, they have to coexist and in most cases cooperate within the wider scope of 5G cognitive management, without the need for a time-consuming human engineering intervention or the use of predefined action rules~\cite{jiang2016machine}. This is also supported in the work~\cite{chen2017machine}, which discusses the solution of a wide variety of wireless networking problems by deploying \ac{ml} approaches.

\label{desiredOver}
A  5G \ac{cnsm} system can deploy appropriate \ml algorithms against two main causes for increasing runtime costs: unnecessary slice resource overprovisioning and the lack of desirable overprovisioning.
While unnecessary slice overprovisioning  regards dedicating more resources than needed (e.g., due to lacking an accurate demand model or due to the inability to predict outstanding demand fluctuations), ``\textit{desired}'' overprovisioning needs to be clarified. 
First off, it prevents resource underutilization and may further enhance \acp{no}' revenue in cases of \acp{sla} that allow to assign to slices virtually more resources than the available physical ones. 
Second, it helps to avoid \ac{sla} violations and corresponding penalties, which are particularly high for critical slices like \eH,  by taking timely overprovisioning actions. These actions enable to correspond quickly to predicted resource need increases such as when an ambulance is moving fast by quickly changing \ac{gnb} on its way to the hospital.
    
Past work in the literature has identified the need for a well-defined and organized way to manage 5G sliced resources intelligently. The authors of~\cite{yahia17} propose a cognitive management architecture with \ac{ml} techniques following the \acf{MAPEk}~\cite{MAPEK2015} control loop. 
Along the same lines, the work of~\cite{nurminenM18} introduced an architectural framework based on the \ac{MAPEk} loop for 5G network management tools to address requirements such as batch, real-time and joint batch plus real-time.
Another interesting work presented in~\cite{luong18} demonstrates how to empower self-organizing networks based on an \ac{ml} model used to cluster and forecast the network traffic of cells.

The current work takes a different approach and tries to complement efforts such as the ones above by covering a gap of a \textit{methodological} approach that systematizes, organizes and automates a series of necessary steps for building efficient \ml models as components of a unified 5G \ac{cnsm} system. 
In essence, we try to design a framework on (i) designing ML models that (ii) can work together in order to identify and apply an appropriate resource provisioning model adaptable to the high network dynamics and demand uncertainties in 5G.

\begin{figure}[t]
   \centering
   \includegraphics[width=0.5\textwidth]{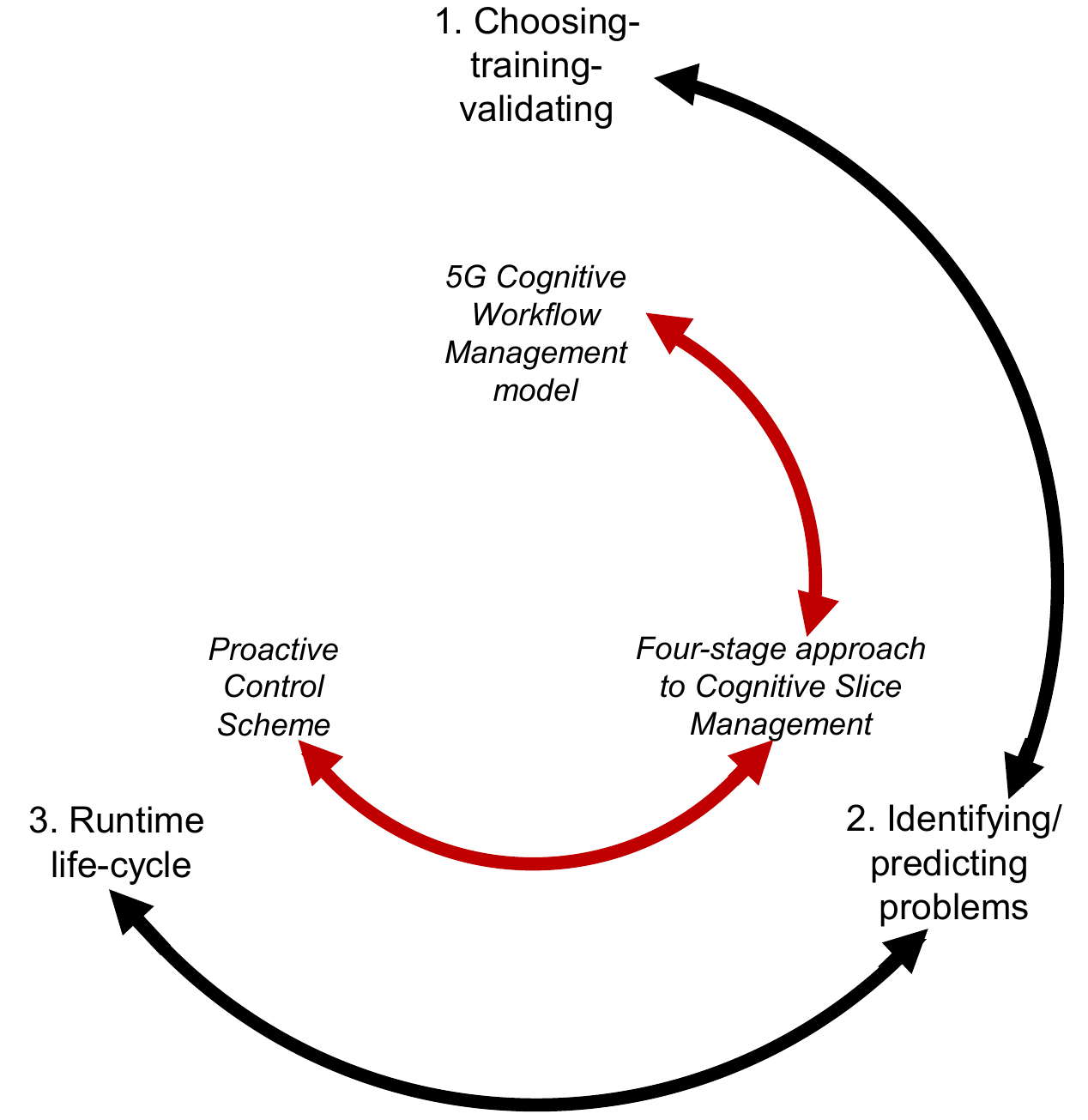}
    \caption{The outer (black) semi-cycle includes the three dimensions, each of which is addressed by a corresponding component of our methodology in the inner (red) semi-cycle. Notice the arrows denoting the coupling between the dimensions. Likewise, there are arrows between the methodology components that show their inter-relation and how they integrate into a unified methodology.}
    \label{fig:3dims}
\end{figure}

Towards such a methodology, we identify three strongly-coupled dimensions (see Figure~\ref{fig:3dims}): 
\begin{enumerate}
	\item \textbf{Choosing--training--validating:} A dimension of \textit{choosing, training} and \textit{validating} an appropriate \ml algorithm targeting a specific problem type.

	\item \textbf{Identifying/predicting problems:} A dimension for \textit{predicting} and \textit{identifying} the exact nature a problem after input from network and slice-logical ``sensor readings''.
	
	\item \textbf{Runtime life-cycle:} A fully fledged \textit{management life-cycle} that defines the runtime \textit{cooperation} of the different \ml models that together form a unified 5G cognitive management system.
\end{enumerate}

Accordingly, the current paper poses a novel Integrated Methodology approach to \ac{cnsm} in virtualized multi-tenant 5G networks. The methodology is built bottom-up, aligned with a concrete eHealth \ucase by keeping a specific example (5G connected ambulance) in mind along with references to other examples and further \ucase scenarios. However, it is elaborated towards deriving a \textit{generic} methodology approach that can cover a plethora of other 5G cognitive slice management use-cases, adding further components as needed. 

\subsection{Contribution}
In a nutshell, the main points of contribution of our methodology are:
\begin{itemize}
	\item 
        \textbf{5G Cognitive Workflow Management model\footnote{For brevity, also simply referred as the ``Workflow''.}:} We adopt a Workflow of four phases that serves as a \textit{step-wise guideline} for abstracting every important aspect for successfully building an appropriate \ac{ml} model, spanning from specifying the exact type of a problem related to 5G \ac{cnsm} up to training, deploying and maintaining the model. This component tries to address the previously identified ``choosing--training--validating'' dimension. 
    \item 
        \textbf{\acf{fcsm}:} We identify four stages to managing 5G slices in a cognitive way that span from gathering slice ``sensor'' readings and identifying a potentially ``anomalous'' situation, up to taking action(s) against it. Note that each stage follows the Workflow internally. This component tries to address the ``Identifying/predicting problems'' dimension.
    \item
      \textbf{Proactive Control Scheme (PCS):} PCS models a fully-fledged runtime approach for addressing arising problems, and a complete life-cycle for 5G \ac{cnsm}. PCS describes the runtime execution of the deployed \ac{ml} models, hence incarnating the deployment phase of the Workflow model. Moreover, it organizes the functioning, the update actions, the cooperation and --if needed-- the redeployment or replacement of running \ac{ml} models with other ones trained specifically for different scenarios. As it becomes clear, \pcs tries to address the ``runtime life-cycle'' dimension.
\end {itemize}

As portrayed in Figure~\ref{fig:3dims}, the above components integrate into a unified methodology, and of which address a corresponding methodology dimension. The Workflow abstracts the internal process of each step in the \ac{fcsm}, while the \fcsm guides the steps for addressing runtime slicing issues by \ac{pcs}, emphasizing on anomaly detection and prediction. Therefore, the three methodology components are intertwined with \fcsm holding a ``linking role'' between the Workflow and \acs{pcs}.

Last, as a by-product of our methodology's practical assessment, we contribute raw and processed 5G \ac{ran} monitoring data~\cite{crawdad:elasticmon2019} from the MAC, RRC and PDCP layers to the Community Resource for Archiving Wireless Data At Dartmouth (CRAWDAD). To our knowledge, this is a rare contribution of \textit{real} 5G \ac{ran} monitoring data to the 5G community.

\subsection{Article structure}

This article is organized as follows. 
Section~\ref{sec:background} introduces the reader to fundamental concepts such as \acp{sla} and the \eH use case as a reference for building the methodology. Moreover, it discusses important related works in the literature regarding the use of \ac{ml} in 5G.
We proceed with posing and analyzing all the components that make up our integrated methodology in Section~\ref{sec:methodology}, and discuss its applicability beyond the scope of the \eH use case.
In Section~\ref{sec:poc}, we demonstrate the merits of the proposed methodology in a practical assessment based on two different \ucase scenarios. 
Finally, we wrap up and outline our future work goals in Section~\ref{conclusion}.

\section{Background and Related Work}
\label{sec:background}
This section introduces the reader to the practical problems faced by \acp{no} while trying to respect their \acp{sla} with 5G slice owners. As a first step, we provide a high-level description of the problem of keeping \acp{no} in line with their \acp{sla} by continuously optimizing network resource allocation to address the high network dynamics in 5G. Second, we outline the role of cognition in respecting \acp{sla}.

Next, we discuss use cases in 5G. We focus first on the \eH use case defined within the context of the EU-initiative H2020 SliceNet project.
This discussion is fundamental, as we take a use case-driven bottom-up design approach based on \eH  to derive our methodology. 
Putting things within a greater perspective, we also discuss two other sample use cases in 5G, namely a strike sample use case and another one on sudden and important events.
Note that we return and explain how our approach can be applied to other 5G use cases after presenting our methodology in Section~\ref{sec:beyondEHealth}. 

Finally, we outline significant works from literature related to how \ac{ml} can be used in 5G.

\subsection{Service level agreements in 5G}
\label{sec:sla}

The heart of the 5G cognitive slice management problem lies in the continuous effort of \acp{no} to respect their \acfp{sla} with slice owners. 
Although there is no general ``recipe''  for \ac{sla} contracts, they are usually along the lines of guaranteeing a particular level of \ac{qos}. 
For instance, giving at least a 2~Mbps download capacity to the \acp{ue} of a 5G slice in 95\% of the times.   
\ac{sla} violations translate directly to \textit{monetary penalty costs} and can even lead to significant revenue losses due to dissatisfied slice owners that can shut down their slices and move them to a competitor \ac{no}.

Unless dictated by an \ac{sla}, the current trend is for NOs to depart away from the obsolete and inefficient model of overprovisioning of the resources, and rather to adapt a \textit{resource multiplexing} model for hosting and controlling 5G slices~\cite{foukas2016flexran, foukas2017orion, EURECOMRuntime}.
Nevertheless, resource multiplexing can put at stake the \acp{sla} of different 5G slices running in parallel and competing for the same physical resources \cite{ksentini2018providing}. 
Consequently, NOs need to timely predict and address possible \ac{sla} violations via continuously monitoring network conditions. 
Some examples of network conditions include network statistics like traffic congestion in the backhaul (BH), the fronthaul (FH) or the wireless part of RAN slice,
throughput and throughput jittering or latency perceived by \acp{ue}. 
Other examples that refer to monitoring network states include the quality of \ac{ue} connectivity services such as the percentage of cellular attachment failures or cellular handover failures. 
Last, monitoring can include clearly user-centric metrics like \ac{ue} battery consumption, which is of particular importance in IoT 5G slicing scenarios.

\subsection{The role of Cognition in respecting Service Level Agreements}
\label{sec:cog-sla}

No matter how quick in predicting and responding \ac{ml}-based solutions can be, still, there are scenarios in which it is inevitable to violate one or more slice \acp{sla}.
Therefore, pragmatically efficient solutions must imply cognitive actions that re-allocate resources among the hosted slices in a way that \textit{optimizes} a \textit{feasible} resource utilization by minimizing the induced \textit{monetary cost} penalties for the NO.
Such actions by an \ac{ml} model could use the cost of the current or predicted utilization of physical operator resources as input.
Resources imply their own type of a cost based on some cost function implied by the \ac{sla}.   

Finally, the cognitive model must avoid wasting resources due to slice \textit{overprovisioning}~\cite{bega2019deepcog} (i) prevent resource-wasting bottlenecks in either the \ac{cn} or \ac{ran}, and to (ii) conform to slice \ac{sla} requirements \cite{chen2018multi} during a near-by future time-window.
Another example refers to expanding \ac{ran} and wireless resources for a particular slice so that the latter can achieve its  \textit{mobility management} and/or wireless \textit{scheduling} goals that derive from its \ac{sla} requirements.

\subsection{Use Cases}
\label{sec:all-usecase}

\subsubsection{The eHealth Use Case}
\label{sec:ehealth}
\label{sec:baseline-usecase}

We consider an eHealth \ucase organized around the concept of a ``5G connected ambulance'' that acts as a mobile edge or hub for emergency medical equipment and wearable sensory devices, enabling to store and to stream \textit{real-time} video of patients' data to the awaiting emergency medical professionals at the destination hospital. Real-time streaming video enables the awaiting professionals to remotely monitor the patient and possibly to provide life-saving remote assistance while in the ambulance under the supervision of the remote specialists with access to sensory and video data. 

The \ac{sla} requirements in the eHealth \ucase are highly demanding, mainly \wrt two aspects:
\begin{enumerate} 
  \item  There is a need to support \textit{\ac{hd}} and  \textit{ultra \ac{hd}} \textit{video streaming} from the 5G connected ambulance to the remote site where professionals reside to serve serious medical emergencies requiring a detailed video made available to remote professionals, e.g. for a serious injury or a stroke. Note, that under extremely stressful network conditions or when an ambulance has to go through a poorly covered area, the \ac{hd} requirement may be degraded to \ac{sd} video quality. 
   
   \item This enhanced and interactive communication between the medical professionals and the remote paramedics requires the \textit{continuous} and \textit{uninterrupted} collection and streaming of data, starting from the arrival of the emergency ambulance at the incident point and lasting until the delivery of the patient to the destination hospital. The goal is for all paramedics to have wearable clothing that can provide \textit{real-time video feed} as well as other \textit{sensor-related data} pertaining to the immediate environment. 

\end{enumerate}
Based on the above, it is clear that the ambulance (i) must remain connected to the 5G network throughout its trip to the hospital, while the required (ii) \ac{hd} video \qos must remain guaranteed under difficult conditions, falling back to \ac{sd} only when it \textit{cannot} be done otherwise. 
``Difficult'' conditions refer to traffic jams impacting the distribution of ambulances to cells and, thus, their share of available physical resources; unexpected network flashcrowds and/or background network traffic; and \ac{mie} under catastrophic scenarios such as massive injuries or extreme weather conditions where multiple ambulances \textit{must} be served at any cost, getting the highest possible \ac{qos} even at the cost of degrading/shutting down other slices over the same physical infrastructure\footnote{
    The role of ML is to predict ambulance paths (including radio handovers) and side-traffic demand from other slices. Then, based on a given slice priority parameter, special ML models define how resources are given to slices. Shutting down, rather than degrading, other slices is an extreme action (the last resort).
}. 
Apart from multiplying the needs for network resources in catastrophic scenarios, note that it may be required to serve the ambulances via \acp{gnb} along road routes ``out of the ordinary'' towards the hospital.    

Clearly, \eH slices must be treated with the \textit{highest priority} compared to other slices in order to have enough resources to deliver the service and to respect the \ac{sla}. 
The HD video requirement, in particular, also raises the \qos standards in terms of the needed amount of resources that satisfy a quite hard \qoe expectation by medical professionals, who must have a crystal-clear view of the patients' condition.

As a final remark, we note that the \eH \ucase requirements fall within the general class of \ete slicing requirements.
\ac{e2e} slice users\footnote{
    Our framework makes no assumption regarding the number of users grouped in a slice. This is up to the slice SLA designers and the developed ML algorithms that manage the slices.
}, i.e. medical professional at a fixed point on the one end and paramedics in a mobile environment on the other, must be \textit{satisfied} based on one or more QoE metrics. At the same time, a series of strict QoS levels must be aligned with the SLA between NOs and slice owners.
Ideally, \ac{qoe} levels should be \textit{automatically extracted} from quantified \ac{qos} metrics, rather than asking users to score their opinion. 
The latter can be impractical to perform for medical professionals who must quickly treat one incident after another during an emergency shift. 
This implies the use of further \ac{ml} techniques that use \ac{qos} as part of their input. Nevertheless, this is a problem that is orthogonal to what we discussed here, thus it is left out of the scope of this article.

\subsubsection{Strike use case}
\label{sec:strike-usecase}

Let us assume a simple \ac{sla} according to which all the \acp{ue} of a slice must have a guaranteed access to wireless cellular resources with at least  2~Mbps of throughput. To simplify complexity even more, let us also assume that the \ac{sla} requirement must hold with a 100\% level of guarantee, i.e: (i) \textit{every} \ac{ue} (ii) must \textit{always} have \textit{access} to a \ac{gnb} and (iii) must \textit{always} enjoy the aforementioned \textit{minimum throughput}.

A typical working day involves residential area users turning on their \acp{ue} and starting to roam  to non-residential areas 
or to hotspots within the residential areas (e.g., students going to schools, people that work in the city center, etc.).
These are  observations \textit{in time and space} characterizing a "normal" slice behavior. 
Thus, the hosting \ac{no} expects to have to increase (resp., reduce) the range of its \acp{gnb} or to turn on (resp. shutdown) some \acp{gnb} 
following user dynamics.
Parallel to such evident actions, a series of other actions must be taken to support traffic dynamics such as increasing (resp. reducing) RAN BH/FH resources (e.g., microwave BH connections to \acp{gnb}), all of which map to cost changes for implementing the \ac{sla}.

Problems start to arise under a strike scenario or any other equivalent event. 
This either sudden or scheduled event directly \textit{unsettles} the \textit{ordinary} traffic pattern in many ways.  
For example, users do not use the bus or metro lines, which decreases the expected \ac{ue} usage during commuting because users that drive their own vehicle do not stream video content as they would do when commuting with mass transportation.
Moreover, some users may not even leave home at all.

In such a \ucase scenario, the \ac{no} must take actions in order to continue to respect the \ac{sla}, 
despite the anomaly in\,the\,traffic\,pattern\,demand. 
In\,addition, slice\,users\,may\,be homogeneous such as in the case of a special students slice, which can make the use-case clearly user-centric.
In bottom line, addressing the observed traffic pattern anomaly translates to preventing needless overprovisioning of resources for certain \acp{gnb}, while allocating more resources to other \acp{gnb} for serving users who, e.g. stay at home rather than move.

\subsubsection{Sudden and important events}
\label{sec:other-usecases}

For the sake of a more complete view on possible use-cases and scenarios, here we briefly comment upon sudden, important and/or emergency event scenarios like traffic demand flashcrowds that occur after an accident, police incident or a natural disaster (fires, earthquakes, floods, etc.).
Note that sudden \& important event \ucase scenarios can coexist with the eHealth use case.
Sticking to the \ac{sla} can involve prioritizing slices, shutting down slices (for security), taking resources from other slices and giving them to eHealth, military or police slices, so as to preserve the desired \ac{e2e} \ac{qoe} for users.

\subsection{Related Work}
\label{sec:relatedWork}

The work of~\cite{imranZ14} identifies the high complexity and \ac{opex} in the upcoming 5G era. The authors focus on \acp{son} as a solution, which falls within the concept of \ac{zsm} adopted in our work.
To this end, they propose a comprehensive framework for empowering \acp{son} with big data to address the requirements of 5G. 
Likewise, to some steps and actions in our workflow, they do data characterization and clarify the needed \ac{ml} tools to transform big data into integrable data forms in a \ac{kb}. 

The authors of~\cite{yahia17} propose a cognitive management architecture with \ac{ml} techniques following the \acf{MAPEk}~\cite{MAPEK2015} control loop. The work also presents a prototype instantiation for two \ac{no} use cases using \ac{lstm} and an \ac{ml} framework for real-time accurate bandwidth prediction for mobile users.

Along the same lines with~\cite{yahia17}, the authors of~\cite{nurminenM18} build upon the Lambda Architecture~\cite{marz2015big} by proposing an extended version of it, namely, the Extended Lambda Architecture (ELA). ELA is a generic unified framework solution for new 5G network management tools. It combines together batch and real-time data processing with adaptive \ac{ml} in a simple Monitor-Analyze-Plan-Execute scheme over a shared Knowledge (MAPE-K) loop~\cite{MAPEK2015}. 
Last, this work provides an experimental tool after the ELA architecture, which tries to address the objectives of mobile operators for cell outage management in 5G.

Another interesting work presented in~\cite{luong18} demonstrates how to empower \acp{son} based on an \ac{ml} model for traffic management after clustering and forecasting cellular traffic. Despite the idea being designed for older generations (GSM, 3G, 4G), it remains largely contemporary in 5G due to a large number and heterogeneity of cells, as well as a variety of traffic characteristics per different cell types.

All of these works have an important contribution towards \ac{ml}-based \ac{zsm} in 5G. However, they leave an important gap for systematizing, organizing and automating all the necessary steps and actions for building efficient \ml models as components of a unified 5G \ac{cnsm} system.
This is where our work comes in place to complement the above.

\section{Integrated Methodology for Cognitive 5G Network \& Slice Management}
\label{sec:methodology}
In what follows, we present and analyze our novel integrated methodology approach.
We start by defining the role of the \ac{kb} and its various aspects \wrt to the methodology components. Then, we provide a detailed description of each methodology component by building our concepts upon the \eH \ucase (Section~\ref{sec:ehealth}), noting, however, that the methodology is able to cover a broader context of use cases.

\subsection{The Knowledge Base}
\label{sec:kb}

Due to its continuous interaction with a process of analyzing data in Phase\,2(b) of the Workflow,  an initial \acf{kb} gets created and then continuously updated and optimized with 
    (i) online monitoring data, 
    (ii) processed/complex data such as labeled data along with desired policies, and 
    (iii) feedback data from the deployed model(s). 
The result is a continuously-refined \textit{\ac{kb}} that interacts with the phases, stages and components of the Workflow, the \fcsm and \ac{pcs}, respectively. 
Therefore, the \ac{kb} as a common abstraction component spanning every part of our methodology. 
As it becomes clearer after discussing Workflow's Phase\,4, the \ac{kb} is an ``integration'' point between the training data fed to the \ac{ml} model and the ``knowledge'' learned by the runtime of the \ac{ml} model itself, an aspect which we present in the \acl{pcs} of Section~\ref{sec:pcs}.
Also, the \ac{kb} can be further enriched with logical network data ``sensors'', an aspect which we cover later as the first stage of our \ac{fcsm} (see ``Pre-phase'' on pg.~\pageref{sec:stage1}). 

Finally, there are two dimensions regarding the policies included in the \ac{kb}: 

\begin{enumerate}
   
    \item
    First, there are policies on both raw and augmented/structured data, which dictate how to share data between stakeholders (i.e., slice owners and \acp{no}) via filters that secure privacy restrictions such as for \acp{no} which do not want to share all/parts of their monitoring information and setup/configuration logs with other \acp{no} or even the slices that they host.
    
    \item
    On the other hand, there are 5G policies in the \ac{kb} that are orthogonal to data. Such policies can be used to train \ac{ml} models or to complement their decisions, e.g. by posing handover restrictions between \acp{gnb} or by posing rules and restrictions \wrt hosting 5G slices across different \acp{no}.

\end{enumerate}

\subsection{5G Cognitive Workflow Management model}
  \label{sec:workflowModel}
\begin{figure}[t]
   \centering
   \includegraphics[width=0.7\textwidth]{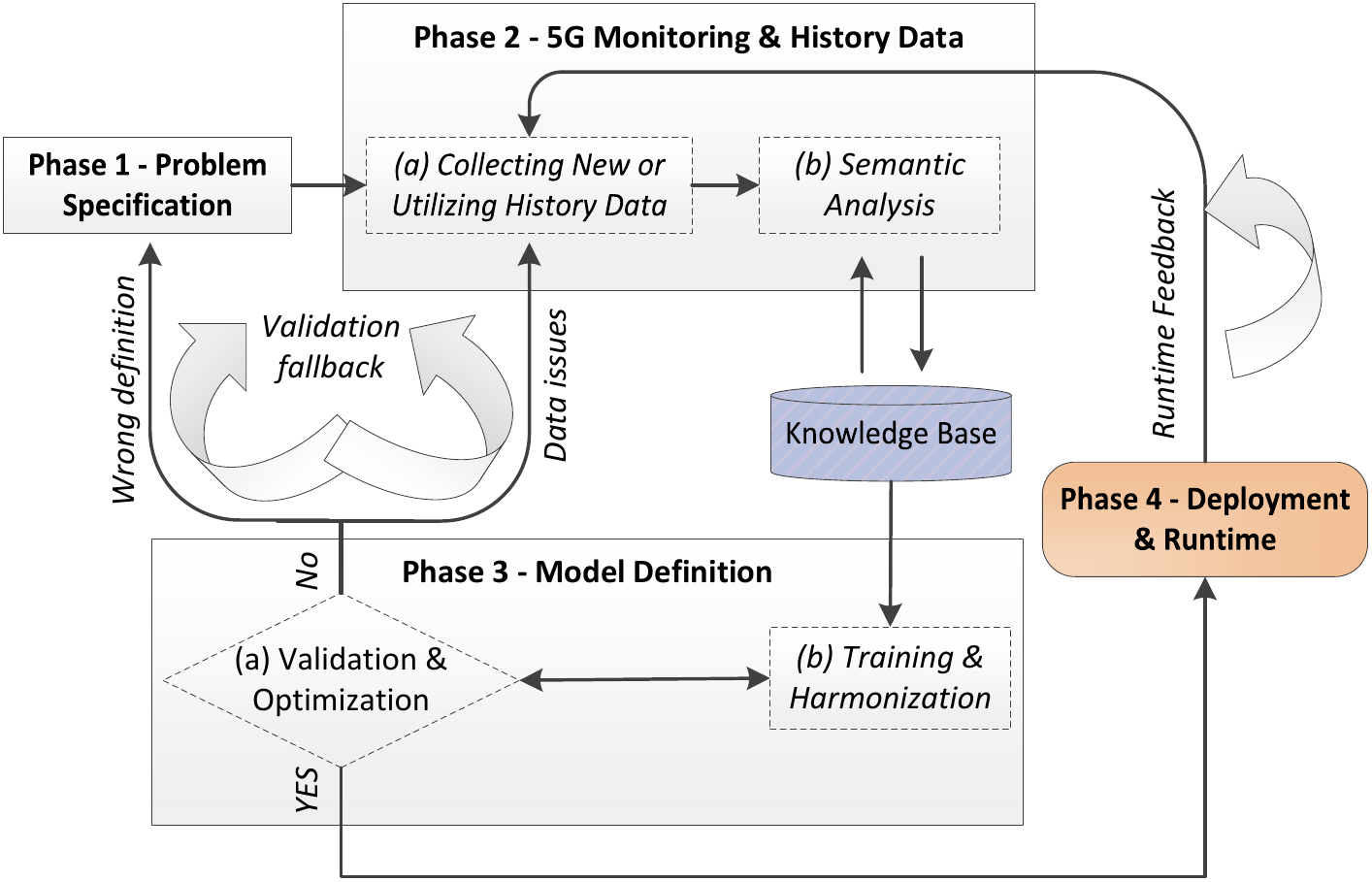}
    \caption{\ac{ml}-based 5G \acl{cnsm} Workflow model. Notice the loop-back to either problem specification or data collection in case the model is not validated.}
    \label{fig:Workflow}
\end{figure}

Figure~\ref{fig:Workflow} portrays the 5G Cognitive Workflow Management model, which is composed of a  step-wise series of four phases including two fallback loops marked as the ``\textit{Validation Fallback}'' and the ``\textit{Runtime feedback}''.
There is also an initial knowledge base built upon history data, which remains neither static nor composed of only raw data, getting eventually evolved to a \acl{kb}.
In what follows, we discuss the \ac{kb} first, and then every phase in detail.

\hfill\break\noindent\textbf{Phase\,1\,-\,Problem Specification:}
\label{sec:problem-specification}

This first phase involves the proper identification of the 5G target problem. Some representative examples of 5G problems in the context of \eH include 
    \textit{anomaly detection} (e.g., the ambulance moves too slow/fast or via areas with poor resource availability); 
    \textit{regression}, \textit{clustering} or \textit{categorization} (e.g., for discovering the statistical relationship between ambulance connectivity and/or route behavior to past performance data);
    \textit{prediction} (e.g., ambulance connectivity will drop after a while or \qos will be degraded); 
    or any other problem from a selection of well known \ac{ml} problem categories.

The consequent two phases and, particularly, Phase\,3 about the model definition on pg.~\pageref{sec:model-definition}, imply high costs \wrt human experts effort, computational load, and the corresponding time cost. 
Therefore, a wrong decision at this stage leads to a significant waste of time and resources until realizing that the problem specification needs to be revisited, forcing to loop back from Phase\,3 to Phase\,1, as denoted by the ``Validation fallback'' loop in Fig.~\ref{fig:Workflow}. 
Not only that but also an added time cost has to be suffered for revising the \ac{ml} model in an effort to address an unsatisfactory learning performance.
Last, a revision to the problem specification and the \ac{ml} model implies a further high cost for repeating the process of selecting the types and amount of raw and/or processed training data needed to train again the revised model (see Phase\,2 next).

\hfill\break\noindent\textbf{Phase\,2\,-\,5G monitoring \& History Data:}
\label{sec:phase2}

This phase embodies to strongly coupled sub-phases: first, for collecting new (raw) data from 5G sources such the \ac{ran} or the 5G \ac{cn}, or for utilizing existing history data from past 5G recordings; followed by a process of analyzing the former input data into semantic information (i.e., useful/meaningful/valuable processed data) to be integrated to the \ac{kb}.

    \hfill\break\hangindent=0.35cm\emph{(a) Collecting New or Utilizing History Data:}\label{sec:collecting-data}
    This sub-phase implies the use of network traffic traces, network performance logs and/or online statistics from 5G sources.
    Thus, the discussions here as well as later in Section~\ref{sec:pcs} about \pcs are interdependent \wrt collecting both offline and online data, including performance monitoring feedback from the deployed runtime model. 
    Depending on the exact nature of the problem we can utilize data from different network layers such as packet-level or MAC layer flow traces \textit{labeled} with some 5G service or application class.
    Using \ac{ran} throughput as an example, we could use spectral efficiency data and data rate(s).
    Spectral efficiency can itself be the output of raw data processing like \ac{cqi} or rank indicator metrics.
    In a likewise manner, the data rate can be the output of
    considering the number of active users, packet size(s) and rates, and/or cellular bandwidth.
    \\
    \indent As previously mentioned, collecting raw and processed data can be done either
    \textit{(i) offline} or \textit{(ii) online} from different network 5G sensors.
    These ``sensors'' are logical entities specially crafted for monitoring 5G events (telemetry readings, topology-related, etc.), raw or processed networking information like network statistics. 
    Whereas offline data involve history data extracted from a repository, online data collection takes place after model deployment during a \textit{perpetual} data monitoring process parallel to the runtime \ac{ml} model instance.
    Therefore, online data implies the use of \textit{real-time network information}
    used either as input\footnote{
        See the discussion on pg.~\pageref{sec:stage3} on output data \textit{dimensioning} in Stage\,3 of \ac{fcsm}. 
    } 
    or feedback for updating the learning model (see ``Runtime Feedback'' loop in Fig.~\ref{fig:Workflow}). 
    Feeding data back to the \ac{ml} model such as for mobile attachment or handover failures helps to maintain (keep training and improving) the deployed model.
    Note that online collected 5G data can be merged to a ``dump/raw'' repository of history data, or get later processed and used to update the 5G \ac{kb}. 
    \\
    \indent 
    Last, to put things into the perspective of eHealth, this sub-phase covers the need to provide the \ml model that allocates resources to the \eH slice with sanitized essential data or features like 
    the statistical distribution of 5G connected ambulances in space (i.e., per \acp{gnb}) and time (during day or night, weekday or weekend, public holiday or season, etc.). Other key feature examples include the \qoe reported by paramedics and remote medical professionals or the recorded \qos measurements.

    \hfill\break\hangindent=0.35cm\emph{(b) Processing, Structuring \& Semantic Analysis:}\label{sec:analyzing-data}
    Evidently, the sub-phase of Semantic Analysis requires a \textit{problem-specific insight} by combining knowledge from both 5G domain experts and service domain experts, e.g. in the context of the eHealth \ucase for video \ac{qoe} optimization~\cite{jiang2016} and/or for guaranteeing seamless transmission of sensory patient data~\cite{Boyi:2015:medicalIot}.
    Raw data and statistics must be analyzed to extract the required key features from data samples, which will next further enrich the \ac{kb}. 
    This presupposes the pre-processing of raw data stemming from both monitoring and history data repositories, which includes: 
    value normalization and discretization; data sanitization, as needed; 
    detecting and correcting or clearing out corrupt/inaccurate records; possibly replacing absolute timestamps with relative times between 5G data recordings; 
    performing dimensionality reduction, i.e. feature selection and feature extraction; and so forth. 
    As a result, the \ac{kb} gets to also include augmented/structured data.
    \\
    \indent 
     The manipulation of online data can take also place in real-time by special applications running in the 5G monitoring system. These applications can span from simple ones such as for calculating average bandwidth consumption per connected \ac{ue} or slice, to more complex ones such as a \ac{ml} model.

\hfill\break\noindent\textbf{Phase\,3\,-\,Model Definition:}
\label{sec:model-definition}
This step happens in two separate but strongly coupled sub-phases: ``Training \& Harmonization'' and  ``Validation \& Optimization''.
The goal within context of the \eH \ucase is to come up a fine-tuned slice maintenance \ml model that guarantees ambulance connectivity and a desired level of \ac{qos}. 
This, also, involves to both train and to validate a candidate \ml model, where validation refers to the \textit{hard} requirements of \textit{not} falsely recognizing a need for reallocating network resources and to \textit{not} assign resources sub-optimally to the \eH slice.
Notice that the two sub-phases are connected with a double-sided arrow, validation testing may imply the need to step back and repeat Phase\,3(a), and then returning back to 3(b), as many times as needed.

    \hfill\break\hangindent=0.35cm\emph{(a) Training \& Harmonization:}
    \label{sec:model-definition-training}
    As briefly mentioned on Sec~\ref{sec:analyzing-data}, it is necessary to undergo a (most often time-consuming) \textit{offline training} step based on analyzed history data and policies (e.g., \ac{no} policies on resource allocation priorities between slices, or slice \qos/\qoe requirements) from the 5G \ac{kb}, which  yields an initial model. 
    In addition, a painful \textit{harmonization} takes place for model parameter tuning. 
    This involves human effort based on accumulated \ac{ml} training experience and \wrt the particular \ac{ml} model, the level of understanding of the problem and of the input data. 
    As a result, parameter harmonization may involve searching in a large space in seek for a ``good'' setup parameter approximation.

    \hfill\break\hangindent=0.35cm\emph{(b) Validation \& Optimization:}
    \label{sec:model-definition-valdation}
    Offline validation is indispensable~\cite{wang2017} in order to evaluate whether the candidate \ml model works sufficiently after its initial learning phase. 
    It involves testing the output initial model to understand if it avoids \textit{over-fitting} and \textit{under-fitting}, both of which lead to poor performance during the final ``Deployment \& Runtime'' Phase. 
    Over-fitting, on the one hand, adapts ``too much'' to the details and noise in the training data, hence the model has ultimately adapted its decisions to noise and outlier data that prevents its ability to generalize to new data during runtime.
    On the other extreme, under-fitting causes the model to neither to fit training data nor to generalize.
    For example, when dealing with a sparse city area in a real-time deployment scenario, training data from dense city areas should not trigger actions analogous to the ones aimed for a massive emergency accident. 
    And vice-versa, data from a sparse area should \textit{not} lead to ignoring an emergency. 
    The earlier would waste resources (hence, harm other, less-prioritized 5G slices); whereas the latter would fail to provide ambulances with guaranteed connectivity,\ac{hd} video streaming and sensory data transmission during a massive emergency event that involves multiple ambulances in an otherwise sparsely inhabited area.
    \\
    \indent 
    Based on the above, a candidate \ml model can be optimized in the sense of lowering model complexity to tackle over-fitting, altering the data volumes fed to the model and examining wrong samples to discover flaws in the model and/or the \ac{kb}. 
    In addition, different parts of the \ac{kb} can be utilized to train different versions of the same or another \ml algorithm in order to tackle the different needs per \ucase scenario (e.g., the spare versus dense area \eH slice above). 
    \\
    \indent 
    Finally, if the model fails to pass the validation sub-phase due to over-fitting, under-fitting or use of corrupt/inaccurate data, then the flow falls back to repeating Phase\,2. Likewise, if the model fails to pass the validation sub-phase because of a wrong problem definition and corresponding model selection,  then the flow needs to fall back to  problem model Phase\,1. We refer to the former as the \textit{Validation Fallback}.
    Also, we remind the reader that if Phase\,2(a) concludes that there is a need to redefine the model by re-harmonizing its parameters and re-training, then there is an internal step back to  repeat Phase\,3(a) and then returning back to 3(b), as many times as needed.

\hfill\break\noindent\textbf{Phase\,4\,-\,5G Deployment \& Runtime Model:}
\label{sec:deployment-runtimel}
At first, a model may work in a best-effort way. 
Assessing its performance in practice involves tradeoff decisions that depend on the nature of the problem. 
Having the 5G networking and slicing concepts in mind, as well as the \eH \ucase, this refers to the cost of resources (e.g., \ac{gnb} energy consumption), accuracy versus overhead and response times (e.g., for ambulance attachment), and the frequency of handovers. 

Given our 5G slicing context, we also include a Runtime Feedback loop from this runtime phase to the \ac{kb} via the steps of Phase\,2. The \ac{ml} model takes real-time input like the current distribution of ambulances to \acp{gnb}, their scheduled routes towards incident sites and the current \qos demand for video streaming resources (HD or SD). 
Then, it retrains itself and yields (i) online \ac{ml} model decisions based on which runtime phase provides (ii) performance output for enriching the \ac{kb}.

\subsection{Four-stage Cognitive Slice Management Approach}
  \label{sec:fourStage}
\begin{figure}[t]
   \centering
   \includegraphics[scale=.7]{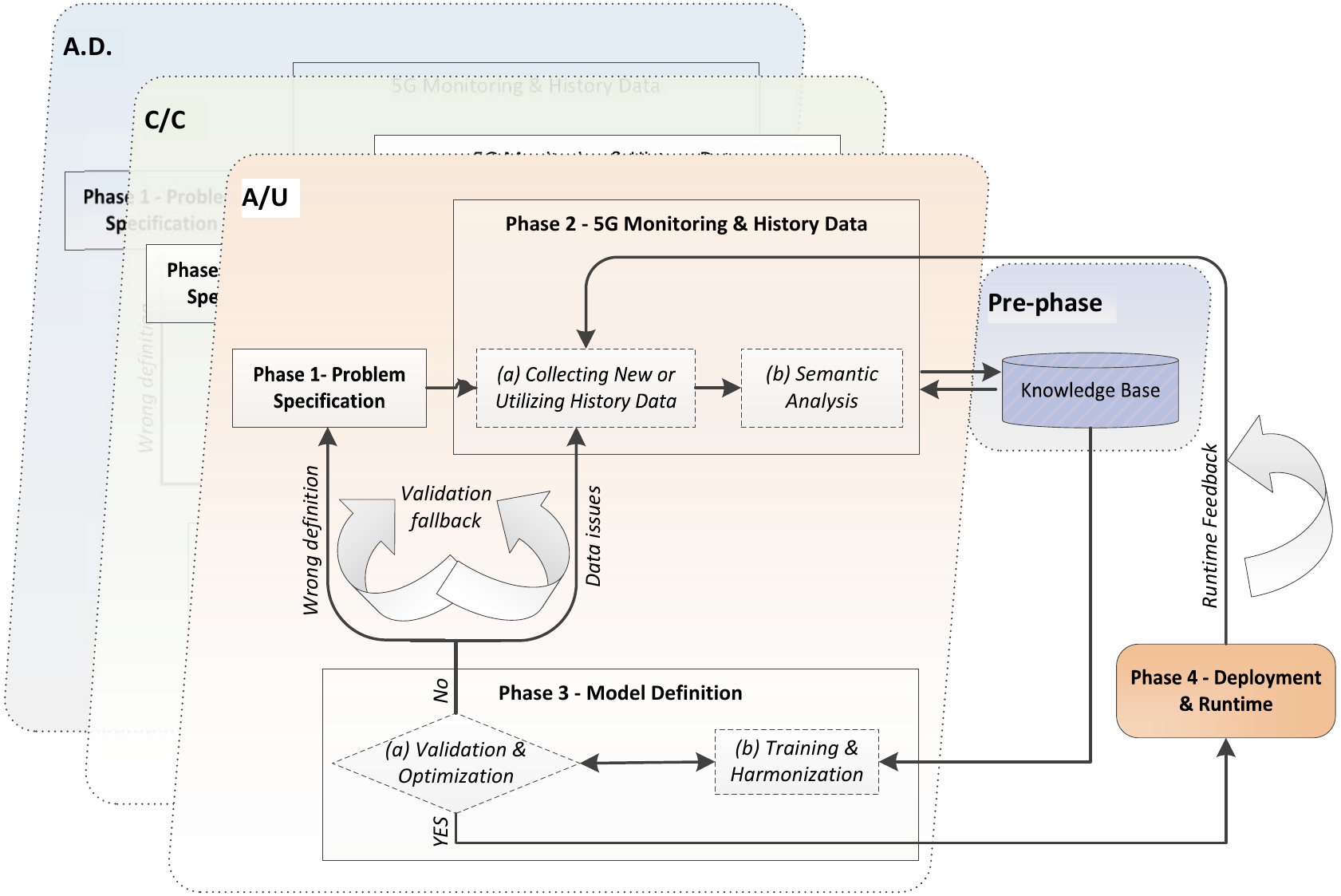}
    \caption{\fcsm: (i)~Pre-phase; (ii)~Anomaly Detection\,(A.D.); (iii)~Clustering and Categorization\,(C/C); and (iv)~Actions and Update model\,(A/U).
    The Pre-phase is operated by both slice owners and underlying \acp{no}, and has a special preparation role for A.D., C/C and A/U that follow after.
    The rest of the stages are operated \textit{only} by \acp{no}, and are internally structured in line with the 5G Workflow of Fig.\,\ref{fig:Workflow}. Message flows between the stages can happen via the KB.}
    \label{fig:4step}
\end{figure}

Figure~\ref{fig:4step} portrays a \acf{fcsm}, as part of our integrated methodology. \fcsm is composed of a 
    (i)~\textit{Pre-phase} for measuring logical ``sensor readings'' from slices;
a stage of 
    (ii)~\emph{Anomaly Detection} based on input parameters, which may signify the need to address a situation ``out of the ordinary''; 
a stage of 
    (iii)~\emph{Clustering or Categorization} to identify the exact nature of the anomaly; 
and a final stage of taking appropriate 
    (iv)~\emph{Actions} and (potentially) \emph{Updating} the \ac{ml} model.
Notice that the depicted \ac{kb} in the Pre-phase is shared among stages.  
Also, the Workflow model of Section~\ref{sec:workflowModel} is internally followed by the later three stages of \ac{fcsm}, which in turn pose the basis of the components of the \acf{pcs} to be discussed latter in Section~\ref{sec:pcs}. 
Notice that the KB in the Pre-phase is later on shared among the consequent stages, as the pre-phase holds a preparation role for A/U, C/C, and A.D. by collecting and placing network measurements in the KB. The 4-CSM does \textit{not} imply direct message flows between the stages. However, there is an implicit exchange of input/output data between the stages via the KB. 
Finally, we note that stages A/U, C/C, and A.D. share the same internal structural template presented in the Workflow model of Section~\ref{sec:workflowModel}, and that these stages pose in turn the basis of all components in the PCS discussed in Section~\ref{sec:pcs}.

\subsubsection{Pre-phase}\label{sec:stage1}

A series of network measurements are collected via logical network ``sensors'' and placed to the \ac{kb}. 
Such measurements can refer to (traces of) \textit{raw} data like the description of an ambulance network attachment attempt at a certain point in time and space along with a failure/success bit; or they can refer to processed network information such as the percent of connection failures per slice, \ac{gnb} region, etc.
The ultimate goal of \textit{slice monitoring} is to gather data about the current conditions in a slice and to place them in a repository as history data. 
As we analyzed in sub-phase 2(a) of our Workflow, 
the purpose here is to provide the necessary input for \textit{learning normal system behavior} via some \ac{ml} model, yet under the series of important considerations in the context of 5G slicing.
Note that the questions posed below as well as the discussion that follows cover how our framework's 4-CSM can be applied under both single- and multi-tenant 5G network scenarios, without a need to adopt any changes in the framework:
\begin{enumerate}
	\item \textit{Who owns training data?} 
	    \acp{no} need access to \ac{qos} data, the \textit{ownership} of which, however, belongs to slices. Likewise, \qoe information can be valuable for training \ac{no}-run models, yet again owned by slices. 
	    Also, \ml algorithms may need  combined input information with sensitive data such as patients' medical background, etc.

	\item \textit{Who pays the cost of resources to collect network data}? 
        Sensors imply a significant system load for resource telemetry and traffic capturing (e.g., sFlow\footnote{
        		http://www.sflow.org/sFlowOverview.pdf
        		}).
        
   \item \textit{How} and \textit{where} is data \textit{stored?}

   \item \textit{How} is data processed?

   \item \textit{Who} manages data? What are the potential implications to \textit{security} and \textit{privacy} for sensitive patients' medical data?
    
\end{enumerate}

Besides the above, \textit{complexity} raises further in \ete slicing scenarios, as 5G slices can be composed of resources from \textit{multiple} \acp{no}, of different types, using (un)licensed radio bands, etc.  
Fortunately, the adaptation of \ac{sdn}/\acp{vnf} allows each \ac{no} to expose its own \ac{vnf} to the slice owner who can in turn use the \acp{vnf} 
to exchange signals with the \acp{no}. 
Such ``signals'' can be either raw data or processed statistics.
This concept is adapted via \acp{vna}~\cite{nikaein2015network, katsalis2017network} in response to an \ac{sla}, leaving room for interaction between \acp{no} and slices, with mutual benefits: efficient resource multiplexing from the \acp{no}' perspective and desired runtime conditions for slices.
For example, specially crafted \acp{vnf} can allow an \eH slice  to have the highest possible priority over less critical or best-effort slices.

Regarding security and privacy, using user mobility tracking data as our example, it becomes clear that slices cannot be expected to willingly share such sensitive data with \acp{no}.
\eH slices, as well as public security slices (police, military, fire department, etc.), would probably \textit{not} trust \acp{no} to share where their vehicles can be statistically found more often.
Existing literature solutions like Bloom filter data structures~\textit{Bloom filters}~\cite{broder2004networkBloom} enable slices to, e.g., pass information about \acp{ue} (ambulances) 
handed over to another \ac{gnb} without exposing their ID. 
Particularly \textit{counting filters}  can be used to multiplex \ac{ue} routes or higher-level information such as resource needs per route.

\subsubsection{Anomaly Detection (A.D.)}\label{sec:stage2} 

Anomaly detection lies at the heart of the cognitive management of 5G slices. 
It is based on input parameters from the pre-phase such as the use of multi-signal input reading,
which are fed to a model that responds with an \textit{anomaly score} output for \textit{short-term prediction}. 
For example, the score can be a difference-based metric that refers to the next time-window, computed after the algorithmically predicted data against the observed data.
\ac{lstm} is a good example of an anomaly detection model that can be combined to recurrent 
\acp{ann}~\cite{donahue2015longlstm} to capture network dynamics. 
For example, in \eH \ac{lstm} can detect an arising anomaly in the routes taken by ambulances or their demanded \qos level. 
Irrespective of the \ac{ml} model, it has to to be continuously updated after each time-window.
The input must refer to data from the current and/or a recent window to predict an anomaly in the next window. 
We return and discuss model retraining on Section~\ref{sec:deployment-runtimel}.

\subsubsection{Clustering/Categorization (C/C)}
\label{sec:stage3}

This is a bridging stage between anomaly detection and that of treating the predicted problem. 
Therefore, its purpose is two-fold. 
First, to \textit{interpret} the anomaly so as to select a specially crafted \ac{ml} model (possibly by replacing the running model) to address the issue or to provide information to the running model. 
As previously referred, one example within the context of \eH involves using different resource allocation models for sparse and dense network areas. Another example refers to the use a different model under public emergencies that can lead to even shutting down  other slices. 
The second purpose is to perform \textit{dimension reduction} on data for the purposes of feature selection and extraction. 
Apart from reducing the cost of gathering data and maintaining the \ac{kb}, dimensioning is necessary for passing the appropriate size and kind of input to the \ac{ml} model running at the next stage.
For instance, the \ml model may simply need to learn (or get better trained by learning) from the number of ambulances or the 
frequency of their handovers in \textit{a greater area} of  \acp{gnb},
\textit{rather than per each} \ac{gnb} or a sector of an \ac{gnb}. 

Whereas \textit{dimensioning} is straightforward with known mathematical models, understanding the anomaly and its cause(s) implies to perform either i) \textit{clustering} or \textit{classification}.
The earlier refers to assigning the detected anomaly to an algorithmically discovered \textit{cluster} of anomalies using an algorithm like K-Means. 
The latter refers to identifying with a predefined class of anomalies 
based on a \ac{ml} probabilistic classifier algorithm, e.g. within the context of \eH a massive incident like a building collapse, a natural disaster or a casual incident increase during vacation times.
By definition, clustering needs further interpretation compared to classification. Labeling clusters can be more difficult, even painful, based on \textit{injecting test traffic} in slices and then ``following'' that traffic to label clusters.

\subsubsection{Actions and model Update (A/U)}\label{sec:stage4} 

This stage contains the appropriate actions taken by a \ac{ml}-based cognitive slice management model tailored specifically for addressing the problem stemming from the previously identified type of anomaly.
The result of these actions is a change in the performance of the slice, e.g. more ambulances can be supported with \ac{hd} video \qos, causing training feedback to the running model. In essence, the model is continuously being updated with performance feedback as a result of its own actions. 
More details are provided in our \pcs model discussion.

\subsection{Proactive Control Scheme}
  \label{sec:pcs}
  
Figure \ref{fig:pcs} portrays our proposed \acf{pcs} as a graph of perpetually interacting components during the runtime of a fully fledged \ac{cnsm} system. Each component has a specific role. 
Note that the context of this scheme is not bound only to models for addressing an already identified anomaly like an upcoming congestion in the \ac{cn} or the \ac{ran} of a slice; but it can also apply to anomaly detection itself with the use of multiple \ac{ml} techniques.

\subsubsection{Monitor \& State}
The monitoring component continuously queries the current \textit{state} in order to extract raw data about the slice such as ambulance to \ac{gnb} link quality based on \ac{rss} or \ac{bh}/\ac{fh} conditions like bandwidth and delay.
``Monitor'' can compute and submit processed data to ``Predict'' in the form of network statistics, e.g. average throughput per user or aggregated per cell.
The monitoring level can be (re)configured to produce more (resp. less) stats or detailed/frequent stats, depending on the needs of the cognition loop and with a corresponding monitoring cost increase (resp. decrease).

\subsubsection{Predict}

This component feeds predictions to ``Action'' that depend on the current monitoring and status, as well as input about slice behavior under normal conditions. 
For instance, an \ml algorithm can learn normal behavior, hence predict the expected load of ambulances, their location and the needed amount of resources based on, e.g., \ac{sla}-tailored re-enforcement learning.  
Example predictions include average throughput per ambulance or per cell during some next time window. 
Additionally, ``Predict'' cluster or categorize anomalies so that appropriate actions can be decided by ``Action''.

Regarding \eH in particular, \pcs must \textit{reserve} some resources in advance, which cannot be allocated fast enough upon demand. This implies giving the highest priority to allocating resources for the 5G connected ambulances even at the cost other traffic. 
\begin{figure}[h]
   \centering
   \includegraphics[scale=.6]{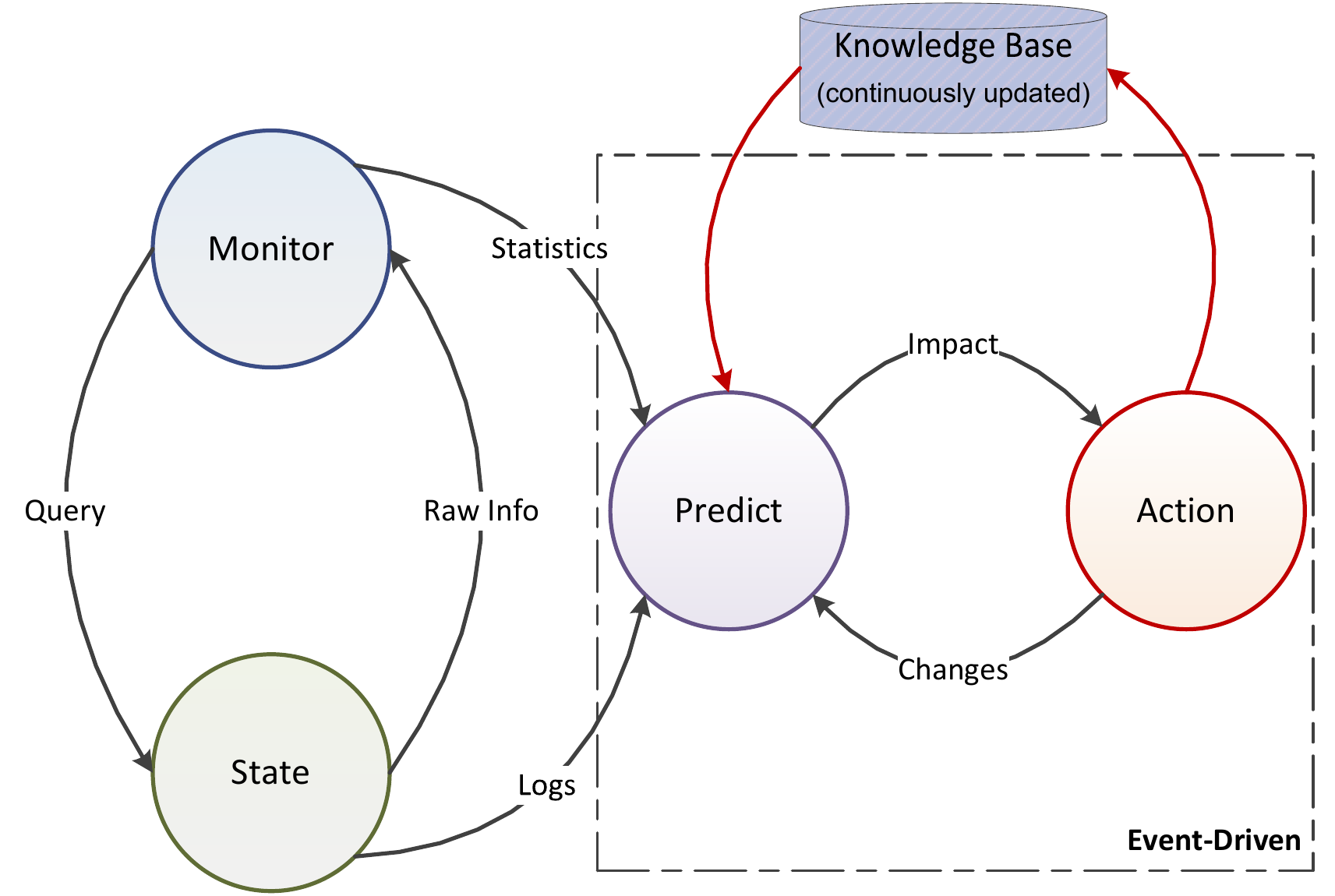}
    \caption{\acl{pcs}. The scheme represents the runtime version of all Workflow phases and particularly phase 5. Likewise, it represents the runtime of 4-SCA with steps like the pre-phase being incarnated by the ``State'' and ``Monitor'', or Anomaly Detection being identified with every component apart from ``Action''. Last, the steps of Clustering/Categorization and Action/Update are identified more with the ``Action'' component.}
    \label{fig:pcs}
\end{figure}

\subsubsection{Action}

Actions in a 5G cognitive management system do not only control the network status.
This component has a dual role as both an internal \textit{(i) \ml control loop} and a \textit{(ii) network control loop} for network actions. 
The \ml control loop can imply actions such as (re)validating a \ml model in use and then replacing one \ac{ml} model with another one.
Other possible actions include instructing to change monitoring level, to (re)start learning, etc. 
Note that all of these actions are \textit{intertwined}, therefore they affect each other. 
Regarding the network control loop, example actions include readjusting transmission power from \acp{gnb}, handing over \acp{ue} to other \ac{gnb}, turning on/off \acp{gnb}, altering the current status of cell breathing (i.e., the range of coverage of \acp{gnb}), adding or reducing \ac{cn} resources (CPU cycles for serving packet queues, bandwidth), changing the \ac{bh} capacity in the \ac{ran}, and so forth.
Notice that some of these actions can affect one slice in particular or even all slices such in the case of altering cell breathing.
Last, the changes caused by actions are fed back to the prediction component along with raw data (e.g. logs) and processed statistics from 
the state and monitor components, respectively, to always update the \ac{ml}-based \ac{kb}.

\subsection{Applicability beyond eHealth}
  \label{sec:beyond-ehealth}

Next, we brief how our methodology is aligned to \eH and how it gets generalized to virtually all use cases in 5G.  

\subsubsection{Aligned with the eHealth use case} 

The root goal is to identify an ordinary functioning pattern of an \eH slice based on sanitized key data or features. 
This allows to build a \ml model for the allocation of HD video resources under ordinary conditions as well as to identify in due time an \textit{anomaly} \wrt the casually exhibited pattern.
An anomaly can be caused by, e.g., a strike that forces ambulances to move via alternative routes (thus, to connect to different \acp{gnb}), or due to, e.g., a massive incident that puts more ambulances in streets and increases the demand for HD video streaming resources.
Pattern anomalies are identified by Anomaly Detection models. Assuming as an example that an unusually high number of ambulances starts to move towards a forest area during a summer period, what follows is that this anomaly is classified by another special \ml model for Clustering/Categorization to a particular type of a problem, in this case a massive incident of a forest fire.
Last, the \ml model trained for (re)allocating resources under normal conditions can be replaced in real-time with a specialized model for emergencies and, more specifically, for emergencies in network areas with a poor infrastructure such as a forest area. In this particular case example, an appropriate model would not only aggressively provision resources in favor of the \eH slice, but may even completely shutdown some other slices.

\subsubsection{Applicability to other 5G use cases}\label{sec:beyondEHealth}

Our methodology can successfully address the needs of 5G use cases beyond eHealth, as its integral parts can be extend with further components as needed. Every Workflow phase is universal to crafting \ml models for \textit{targeted} 5G cognitive management scenarios by (i) treating every \ml problem \textit{differently}, (ii) creating and (iii) maintaining a special \ac{kb}, and by (iv) training and validating a \ml model that is targeted for the needs of a particular slice. Each \ac{fcsm} stage can be applied beyond \eH starting from a common need for ``sensing'' data as a pre-phase to learning \textit{normal} slice \textit{behavior} within the context of any use case. Then, continuing with detecting \textit{pattern anomalies}, which signify a change and a potential problem such as content demand flash-crowds or unusual \ac{ue} movement and/or concentration in an urban environment.  What follows is anomaly detection as well as Clustering/Categorization as cornerstones for cognitive 5G slice management in all use-cases. Last, we remind the reader that \ac{pcs} incarnates the runtime of virtually every 5G use case.

\section{Methodology in Practice}  
\label{sec:poc}
In what follows, we demonstrate the merits of systematizing, organizing and automating all necessary steps and actions in our methodology towards building and deploying efficient \ml models as components of a unified \ac{cnsm} system.
To achieve this, we present a practical assessment using two different \ucase scenarios both of which are related to \ac{wbcqi}. In general, \ac{cqi} is an important4-bit integer metric that denotes how good or bad the communication channel quality is, based on the observed \ac{sinr} that is reported back to the \ac{gnb} by a \ac{ue} to indicate a suitable downlink transmission data rate.
\ac{wbcqi}, in particular, represents an effective \ac{sinr} over the entire channel bandwidth.

Before we proceed with defining the scenarios below, we \textit{stress upon} the fact that we do \textit{not} propose nor intend to produce a realistic \ac{ml} model as part of an integrated \ac{cnsm} system. Our goal is to provide the reader with a \textit{\ac{poc}} evaluation of our methodology via posing two different \ucase problems and exemplifying how it can be necessary to 
    (i) loop back to Phase\,1 in our Workflow for redefining the 5G problem (in Scenario 1), or to
    (ii) loop back to Phase\,2 for repeating sub-phase\,2(a) (in Scenario 2).

Upon finishing with exemplifying the Workflow, we discuss how our trained regression model for the Scenario\,2 can fit in the contexts of the \fcsm and \ac{pcs}, and comment on how it can be combined with other cognitive models in an integrated \ac{cnsm} system.

\subsection{Proof of Concept Use Case Scenarios}

    \emph{Scenario 1\,--\,Estimating \ac{wbcqi} in sleeping \ac{iot} device scenario:} This first scenario regards the ability to maintain the desired \ac{qos} level for an \ac{iot} sensory device when the serving \ac{gnb} has \textit{no} knowledge of the device's contemporary \ac{wbcqi} because the sensory device goes periodically to sleep; thus, when it wakes up it starts to receive data with either old or unknown channel information.

    \emph{Scenario 2\,--\,Estimating \ac{wbcqi}:} This second scenario falls in the context of \eH and regards the ability to maintain the desired \ac{qos} level for a service that allows the doctor to stream his video from the hospital to the ambulance or remotely control medical devices to treat a patient inside a moving ambulance. 
    In \eH it is important to 
    guarantee the seamless mobility of ambulances, which gives priority to frequently tracking \ac{rrc} metrics like \ac{rsrp} and \ac{rsrq} with the highest granularity on a per ms basis. 
    However, \ac{mac} metrics like \ac{wbcqi} can be monitored in the order of tens of ms, as frequent monitoring implies significant costs per slice in 5G, with the amount of monitoring data becoming huge \wrt  using online or for storing offline for future uses. 
    Nonetheless, \ac{wbcqi} is important for inferring information about the downlink quality of the ambulance, hence for maintaining HD video quality or for a critical \ac{urllc} medical remote control service.

\subsection{5G Workflow}

\subsubsection{\textbf{Phase 1; Regression model for wbCQI}}
\label{sec:assess:problem-specification}
    We specify that both scenario problems fall in the well-known category of \textit{regression} analysis for ``predicting'' \ac{wbcqi} as a dependent value from other independent monitoring metric values. A proper \ac{wbcqi} regression  model will become part of our \ac{cnsm} system, being responsible for assessing the contemporary status of \ac{qos} in either of our \ac{poc} \ucase scenarios.

\subsubsection{\textbf{Phase 2; 5G monitoring \& History Data}}
\label{sec:assess:phase2}

    \paragraph{Collecting New Data}
    For the purposes of this \ac{poc}, we collected new and raw 5G \ac{ran} monitoring data using our a prototype version of \textit{ElasticMon v0.1}\footnote{
        https://gitlab.eurecom.fr/mosaic5g/elasticmon/tree/develop.
        For more information on how to access the code and resources, visit the Mosaic5G web site: http://mosaic-5g.io/membership/.}, 
    a novel elastic monitoring 5G framework built over the FlexRAN~\cite{foukas2016flexran} for OAI-RAN and OAI-CN\footnote{
            https://snapcraft.io/oai-ran,~~https://snapcraft.io/oai-cn.
    }. 
    The collected raw data contain over a 100 metric categories from the \ac{mac}, \ac{rrc} and \ac{pdcp} layers as exposed by the FlexRAN controller recorded for 1\,\ac{ue} in a JSON format. 
    These raw data are organized in 5 different raw datasets recorded for one \ac{gnb}, each of which corresponds to one out of 5 different \ac{ue} mobility scenarios by following different motion and distance patterns.

    Note, that we have contributed these raw as well as processed versions of the datasets from Phase\,2(b) to CRAWDAD~\cite{crawdad:elasticmon2019}. To our knowledge, this is still a rare contribution of realistic 5G \ac{ran} monitoring data to the 5G community.

    \paragraph{Processing \& Structuring Analysis}
    It was necessary to process, structure and, in general, to clear up these raw datasets in two steps:
    
    \emph{-- Step\,1;\,Pre-processing:} Pre-processing takes place to sanitize raw recordings and to reduce the number of metrics per measurement from over a 100 to 42. Pre-processing is necessary for a series of reasons:
            \begin{itemize}
                \item \textit{Adding a timestamp:} Exact dates in raw measurements do not give useful information. It is necessary to add timestamps inside the recorded JSON tree of each measurement. This is needed for computing the time elapsed between consecutive measurements.
                
                \item \textit{Cleaning out static values:} Omitting specific metric fields that do not change over time. Such metrics maintain a constant value across measurements regardless of the \ac{ue} being in motion or not. Therefore, they offer no valuable information for prediction. Note that the remaining "dynamic" metrics after this step drops to 42.

                \item \textit{Adjusting corrupt/inaccurate metric values:} There where some measurements with corrupt/inaccurate values. 
                The problem was addressed based on the type of metric and number of consecutive corrupt/inaccurate values in two different ways by replacing evident corrupt/inaccurate values with 
                    (i) the \textit{median} value of their neighboring rows, or 
                    (ii) the \textit{mean} value over a period of time (e.g., past 100\,ms) out of a series of neighboring rows.  
            \end{itemize}
    
    \emph{-- Step\,2;\,Dimensionality reduction \& Feature extraction:} In this step, the number of metric features got further reduced from 42 to 15 according to the following: 
    \begin{itemize}
        \item \textit{Correlation analysis:} We produced a correlation matrix\footnote{
                    Based on DataFrame\_Corr:
                    https://pandas.pydata.org/pandas-docs/stable/reference/api/pandas.DataFrame.corr.html.
                }
            that enabled us to proceed with a corresponding appropriate \textit{feature extraction} based on the level of correlation against \ac{wbcqi}. In further, this process allowed us to reduce the dimension of the training input and testing datasets that would be used next for the model definition phase.
        
        \item \textit{Feature exclusion:} We excluded MAC statistic metrics like ``mcs1Dl'', which are directly calculated based on \ac{wbcqi}, as the purpose of the envisioned regression model is to predict the dependent \ac{wbcqi} out of independent metric values. 
    
    \end{itemize}
   
   The resulting analyzed datasets got ready-to-use for training and testing, being composed of recordings taken in approximately 30000 scenarios.

\subsubsection{\textbf{Phase 3 -- Model Definition}}
\label{sec:assess:model-definition}
We considered the following \textit{candidate} regression model approaches in our effort to find a best-fitting regression model for predicting \ac{wbcqi}:
\begin{itemize}
    \item \textit{\acs{lasso}:} \ac{lasso} is a regression method that penalizes the absolute size of predictor coefficients. It is a good approach when dealing with highly correlated predictors. 
    
    \item \textit{Elastic\,Net:} Elastic\,Net is a linear combination of the penalties of both \ac{lasso} and ridge regression (\aka weight decay). Note that Elastic\,Net encourages a grouping effect among highly correlated predictors, which enables better control of the impact of each predictor on the \ac{wbcqi} prediction.
        
    \item \textit{Random\,forest and XGBoost:} These are tree-based models. A random\,forest uses an ensemble learning method for regression by constructing multiple decision trees at training time and outputting the mean regression of the individual trees. \ac{xgboost} on the other, grows a tree with a ``boosted'' training approach according to which it learns each variable-to-variable relation and grows a tree accordingly.
        
\end{itemize}

    \paragraph{Training \& Harmonization}
    \label{sec:assess:model-definition-training}
    We considered the $x_i^2$, $x_i^3$, $\sqrt x_i$ and $\sqrt[3]x_i$ of the best 15 features $\{x_1, x_2, \cdots x_15\}$ from Phase\,2 alongside those 15 features in our training data frames. 
    This causes a x5 increase in the number of features (from 15 to 75), which adds up to the level of training complexity for all of our candidate models. 
    However, it helps to capture any polynomial relations between the independent feature metrics and \ac{wbcqi}. 
    We chose not to get into the details of harmonizing the parameters of each candidate regression model, as this would imply many details orthogonal to showcasing the steps of our methodology.

    \paragraph{Validation}
    \label{sec:assess:model-definition-validation}        
    We chose to make a scenario-based split, where the training set includes data corresponding to all mobility patterns and the validation set includes its own patterns. By doing so, we aim to provide more insight on \textit{where} and \textit{why} the models fail to predict or become less accurate. Using a k-fold cross validation split, as it is common to do, would have not enabled us to discover the exact data series in time and space (due to the motion pattern) that cause performance to drop. Still, we tried to keep a 90\% to 10\% ratio between training and validation data set sizes, yielding a training set of 26082 data frames and a validation set that consists of 2959 data frames.

    In what follows, we first explain why for Scenario\,1 we need to loop back to Phase\,1. Then, we explain why for Scenario\,2 we need to also loopback, this time to Phase\,2(a). Note that from that point and on, we stop referring to Scenario\,1 for the rest of this \ac{poc} methodology assessment, and continue only with Scenario\,2.
    
    \begin{itemize}
     
        \item\textit{Scenario\,1: Loop back to Phase\,1:} 
        Regarding Scenario\,1, specifying the the problem falls into the category of regression is \textit{wrong}, thus we must fall back to Phase\,1. 
        The reason is that a regression model would need to use \ac{rrc} metrics like \ac{rsrp}, \ac{rsrq} and \ac{phr}. But just like \ac{wbcqi}, these metrics are normally reported back to the \ac{gnb}, which means that when the \ac{iot} device is put on a sleep mode, none of these data would be available as input to the deployed model at runtime. In fact, according to \ac{lasso} the most important 6 fields are $\sqrt{RSRP}$, $RSRQ$, $\sqrt{PHR}$, $\sqrt[3]{RSRP}$, \ac{rsrp} and \ac{phr}.  A similar conclusion applies to the case of the other models too.
       
        \vspace{.25cm}
   
        \item\textit{Scenario\,2: Loopback to Phase\,2(a) and repeat Phase\,3:} 
        We notice a pattern \wrt the highest prediction errors observed for all models. For example, in the case of \ac{lasso}, the highest errors are presented in Table\,\ref{table:1}. Similar errors about actual values "3" are observed with the rest of the models as well. 
        Driven by this, we returned to the training and test sets and observed that there are \ac{wbcqi} values, as continuous series with values such as 14, 13, 10 or 8 where interrupted in parts by a small series of values equal to 3. These value recordings stand easily out as disrupted because of the consistency and accordance of the data series with the mobility pattern of the user in time and space. Based on an expert's view it is impossible for \ac{wbcqi} to suddenly and ephemerally drop to a such a lower value. The latter is not consistent with the specific point in time and space-distance from the \ac{gnb}, in an environment without any interference sources (other \acp{gnb} or \acp{ue}). Note that is a prominent example to remind the reader that the quality of all steps and stages in the process of designing, training and deploying ML models depends on human expertise.   
        
        \vspace{0.125cm} 
        To address this problem, we \textit{fall back to Phase\,2}. We do not need to take actions in Phase\,2(a), yet in Phase\,2(b) we replace the corrupted \ac{wbcqi} values with the \textit{median} value of their neighboring 100 recorded rows, i.e. the preceding 50 ones and the succeeding 50 ones. 
%
%
        \begin{table}[!htp]
            \centering
            \begin{adjustbox}{width=0.35\textwidth,center}   
             \begin{tabular}{c|c|c}
                \textbf{Actual}  & \textbf{Predicted} & \textbf{Error} \\ \hline
                3 & 14.17 & 11.17 \\ \hline
                3 & 14.16 & 11.16 \\ \hline
                3 & 14.19 & 11.19 \\ \hline
                3 & 12.88 & 9.88 \\ \hline
                3 & 7.88 & 4.88 \\ \hline
                6 & 9.82 & 3.82 \\ \hline
             \end{tabular}
            
            \end{adjustbox}    
            \caption{Highest prediction errors for \ac{lasso}.}
            \label{table:1}
        \end{table}

        After repeating the validation, we get the results presented in Table~\ref{table:2}, showing a  performance comparison between all models and a combined model (about which we comment later on) against the testing set. Apart from the prediction accuracy, the table contains the \ac{rmse} and the \ac{mape}.
        As denoted by its name, \ac{rmse} measures the differences between predicted values and actual values, whereas \ac{mape} measures prediction accuracy and can be used as a loss function for our models' performance. 
        Both  \ac{rmse} and  \ac{mape} express average model prediction error and are negatively-oriented scores (i.e., the lower, the better). Their main difference is that \ac{rmse} gives a relatively higher weight to large errors, which important \wrt to large-quality deviations based on \ac{wbcqi}.

\begin{table}[h!]
\centering
\begin{adjustbox}{width=0.55\textwidth,center}   
 \begin{tabular}{r|| ccc}
    \multicolumn{1}{r}{}
    & \textbf{\acs{rmse}}
    & \textbf{\acs{mape}\,(\%)}
    & \textbf{Accuracy\,(\%)} \\ 
 \hline
    Lasso & 0.153 & 10.99 & 76.71 \\ \hline
    ElasticNet & 0.171 & 11.91 & 72.45 \\ \hline
    Random &  &  &   \\ 
    Forest & 0.137 & 6.71 & 83.67 \\ \hline
    XGBoost & 0.125 & 6.3 & 88.34 \\ \hline
    Combined &  &  &  \\ 
    Model & 0.124 & 6.08 & 88.5 \\ \hline
 \end{tabular}

\end{adjustbox}    
\caption{Performance comparison of all models.}

\label{table:2}
\end{table}

\vspace{.125cm}
        All evaluation results in Table\,\ref{table:2} indicate that \ac{xgboost} clearly outperforms the rest of the models. It is more accurate parallel to having less \ac{rmse} and \ac{mape} error scores. 
        Random Forest ranks as a second best option, with \ac{lasso} and Elastic\,Net following next.
        %
        Nevertheless, a closer look to the distribution of errors across all valid \ac{wbcqi} values from 1 to 15 is \textit{not} uniform. 
        This means that there is no single model which has a globally optimal performance across all \ac{wbcqi} index classes. To provide HD video, the choice of the appropriate model has to be based on the ability of the model to be more accurate concerning service/scenario-specific \ac{wbcqi} index values. Alternatively, we can create a model that combines all of the above regression models in seeking for an optimal prediction of \ac{wbcqi}.   

        \vspace{.125cm}
        For this reason, we step back to sub-phase (a) and create, harmonize and retrain a \textit{Combined Model} (see last row in Tables~\ref{table:2}). 
        The Combined Model is comprised of the \textit{weighted average} of all models after trying all possible weight combinations and assessing the resulting performance of each weighted combination against the testing set.
        Specifically, \ac{lasso} does not contribute at all, whereas Elastic\,Net, Random\,Forest and \ac{xgboost} contribute with 1\%, 21\% and 78\%, respectively. Notice that apart from having the highest accuracy, the combined model achieves in principal better error performance figures as well.

        We are now ready to proceed to Phase\,4. We assume that we select to deploy the Combined Model.
    \end{itemize}

\subsubsection{\textbf{Phase 4 -- Deployment}}
\label{sec:assess:deployment}

The Combined Model will be deployed in our system to predict the \ac{wbcqi} of \acp{ue} in real-time. What follows is to describe how this \ac{ml} model fits to the \ac{fcsm} and \ac{pcs} methodology components.

\subsection{Four-stage Cognitive Slice Management}

\subsubsection{Pre-phase}\label{sec:pract-stage1}

This is an important stage in the \fcsm for scenario~2. We remind the reader that the combined regression model needs input such as $RSRP$, $RSRQ$ and $PHR$ among others for the \eH slice. The first question that needs to be answered is \textit{``Who owns training data?''}. Given that these are \ac{rrc} layer metrics it is more likely that their ownership belongs to slices than to \acp{no}, yet access to the measurements or the estimated \ac{wbcqi} is needed by the \acp{no} in order to proceed with the necessary actions for offering the promised \ac{qos} level and respecting the \ac{sla}. 
Regarding \textit{``who pays the cost to collect such network data?''}, recall the possibility of involving more than one \ac{no} at the physical level to guarantee the very strict requirements of this \ucase and to always deliver the \eH slice. This means that measurement collection, as well as the corresponding costs, should be taken over the corresponding \acp{no}, again as part of their obligation to respect the \ac{qos} promised in the \ac{sla}. 

The remaining important questions are subject to the \ac{sla} agreement with the \eH slice, referring specifically to (i)~\textit{"how and where is data stored?"}, (ii)~\textit{"how is data processed?"}, (iii)~\textit{"who manages data?"} and, last, (iv)~\textit{"what are the potential implications to security \& privacy?"} noting for (iv), however, that this does not have privacy implication on sensitive patients' medical data. In fact, it covers \textit{accountability} about for the \ac{qoe}, i.e. the perceived medical care in the ambulance, rather than privacy.

\subsubsection{Anomaly Detection (A.D.) \& Clustering/Categorization (C/C)}
\label{sec:pract-stage2-3}

Our regression model is orthogonal to anomaly detection as well as to clustering and categorization. However, in the context of scenario~2, other models like \ac{lstm} or ``k-nearest neighbors'' can be fed with the same input as our combined regression model and the predicted values by our combined regression model. As a starting point, we can use the features extracted for the explored regression models which compose the combined model, included further or other features as needed.

\subsubsection{Actions and model Update (A/U)}\label{sec:pract-stage4} 

Actions refer to trying to adjust all parameters need after the predicted \ac{wbcqi} such as slice bandwidth to maintain the desired \ac{qos}. Continuous updates of the models underlying the combined model are based on actual \ac{wbcqi} values and predicted ones per composing model. This is more convenient to do offline, as there is more than one underlying model involved in this case.

\subsection{Proactive Control Scheme}\label{sec:pract-pcs}

Finally, our regression model has, clearly, two potential placements in either the ``Predict'' and/or ``Monitor'' components of \acf{pcs} (see Fig.~\ref{fig:pcs}), which depends on a more high-level design approach to \ac{cnsm} for this use case and for capturing scenario~2. To avoid confusion, \ac{wbcqi} prediction by the combined regression model does not strictly refer to the (future) prediction of the \ac{qos} or \ac{qoe}, but rather to the estimation (\aka ``prediction'') of a specific metric from other metric measurements. If \ac{wbcqi} is directly used to take actions, then the combined model underlies ``Predict'', hence continuously feeding with \ac{wbcqi} estimations the ``Action'' component. Alternatively, if the combined regression model is used to estimate \ac{wbcqi} for other models (anomaly prediction or for classifying the current network state), then in such a case it underlies the ``Monitor'' component, feeding with  \ac{wbcqi} estimations the ``Predict'' component. 

\section{Conclusion and Future Work}
\label {conclusion}
We propose a novel unified Methodology approach to \emph{Cognitive} Network \& Slice Management in virtualized multi-tenant 5G networks with the application of \ac{ml}.
Covering a gap in the 5G literature, our methodology eases the complexity of all the necessary actions for crafting and deploying efficient \ml models.
It follows a bottom-up approach that stresses upon the role of \emph{anomaly detection} as a cornerstone for cognitive management, and it is comprised of three intertwined components, namely, 
(i) a \emph{5G Cognitive Workflow} design model, 
(ii) a \emph{\acl{fcsm}} and 
(iii) a \emph{\acl{pcs}}.
Last, we note our contribution~\cite{crawdad:elasticmon2019} with raw as well as processed 5G monitoring datasets to CRAWDAD, as a rare contribution of \textit{real} 5G \ac{ran} monitoring data to the 5G community.

Future work includes applying and testing our methodology in a greater number of realistic \ucase scenarios within and beyond the context of the SlisceNet project, leveraging further real network slice data, \ac{oai}, and our Mosaic-5G\footnote{
    http://mosaic-5g.io/
} constellation of 5G platforms~\cite{2018llmec}.
One direction is to explore \textit{Q-learning} such as in our most recent work of~\cite{arel3p2020}. There, we follow our \ac{cnsm} methodology to study, design, test and deploy an adaptive reinforcement model model as part of a \pcs for VNF placement in a realistic city-wide 5G testbed and use case hosted by the University of Bristol. Such an example exhibits the ability of reinforcement learning and other online models like recurrent \acp{ann} to be developed with our methodology as \ac{pcs} components for controlling end-to-end slices from the \ac{cn} to the \ac{ran} segment.
The goal is to optimize the concurrent allocation of different types of resources in end-to-end and \ac{mec} slicing scenarios, rather than studying the simplified cases of slicing a particular resource type in a particular network segment (e.g., only \ac{ran}) or a ``narrow'' problem like interference control~\cite{galindo2010}.  
In general, \ac{pcs} fits well to the reinforcement learning concept due to the interplay between ``Action'' and ``Predict'' that resembles the actions of a Q-learning ``agent''. 
Finally, we intend to investigate how congestion pricing models such as in~\cite{vasilakos2016niche, vasilakos2017MPM} can enhance reward assessment schemes for Q-learning-based \ac{pcs} algorithms.

\section*{Acknowledgements}
Research and development leading to these results has received funding from the European Union (EU) Framework Programme under Horizon 2020 grant agreement no. 761913 for the SliceNet project and Horizon 2020 grant agreement no. 762057 for the 5G-PICTURE project.

\bibliographystyle{IEEEtran}
\bibliography{main}

\begin{minipage}{0.3\linewidth}
    \includegraphics[width=1.5in, clip,keepaspectratio]{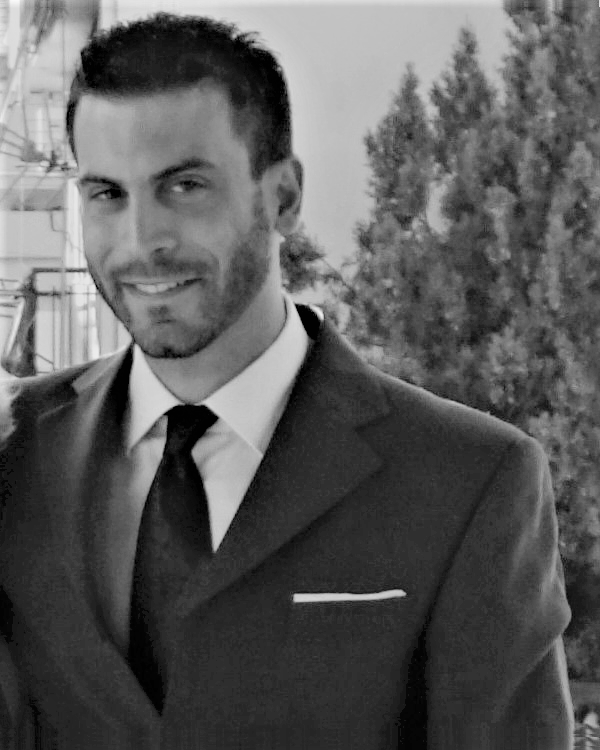}
\end{minipage}\hfil
\begin{minipage}{0.7\linewidth}
    Xenofon Vasilakos received the M.Sc. degree in Parallel and Distributed Computer Systems from Vrije Universiteit Amsterdam, and the Ph.D. degree in informatics from the Athens University of Economics and Business with a focus on Information-Centric Networking architectures, protocols, and distributed solutions. Currently, he is a Research Fellow with the University of Bristol, Bristol, the U.K., where he is a member of the Smart Internet Laboratory and the technical lead researcher of the Zero Downtime Edge Application Mobility (MEC Mobility) project. He has participated in various EU and national funded research projects such as 5GPPP SliceNet and the FIA award-winning FP7 project PURSUIT. His current research interests include 5G/B5G technologies with a focus on Multi-access Edge Computing based on cognition approaches inspired by machine learning models toward self-managed networks. He is also involved in the areas of Internet of Things, Software-Defined Networking, Network Function Virtualization, and network slicing in the context of 5G. Dr. Vasilakos was a recipient of an excellence fellowship grant from the French government (LABoratoires d’EXcellence), and has received an accolade and awards for his academic performance from the Greek State Scholarship Foundation.

CV: http://pages.cs.aueb.gr/\textasciitilde xvas/pdfs/detailedCV.pdf
\end{minipage}

\begin{minipage}{0.3\linewidth}
    \includegraphics[width=1.5in, clip,keepaspectratio]{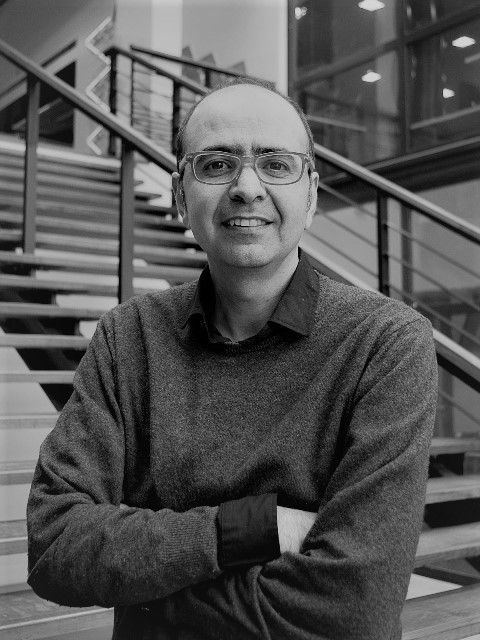}
\end{minipage}\hfil
\begin{minipage}{0.7\linewidth}
    Navid Nikaein is a Professor in the Communication Systems Department at Eurecom. 
He received his Ph.D. degree in communication systems from the Swiss Federal Institute of 
Technology EPFL in 2003. Broadly, his research contributions are in the areas of experimental 
4G-5G system research related to radio access, edge, and core networks with a blend of communication
and computing, and more recently data analysis with a particular focus on realistic use-cases. 
He is a board member of the OpenAirInterface.org software alliance as well as the founder of the 
Mosaic-5G.io initiative whose goal is to provide software-based 4G/5G service delivery platforms.
\end{minipage}

\begin{minipage}{0.3\linewidth}
    \includegraphics[width=1.5in, clip,keepaspectratio]{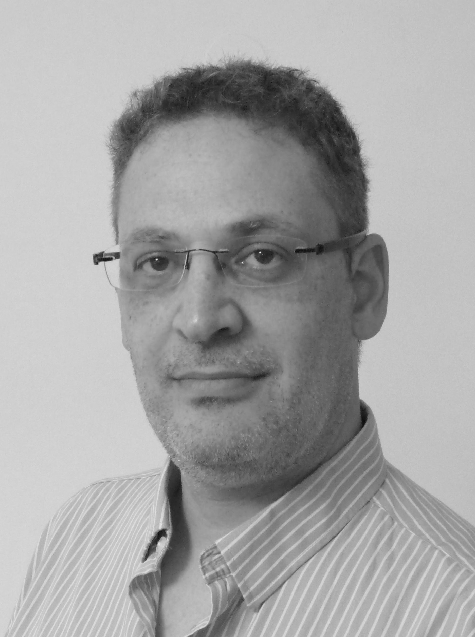}
\end{minipage}\hfil
\begin{minipage}{0.7\linewidth}
    Dr. Dean H. Lorenz received his B.Sc. in Computer Engineering and Ph.D. in Electrical Engineering, both from the Technion -- Israel Institute of Technology. Dr. Lorenz is Researcher at IBM Research -- Haifa, where he is a technical leader in the Cloud Architecture Networking group, in the Hybrid Cloud department. He has more than 20 years of experience in research, hands-on development, and innovation in Networking, Virtualization, Storage, and Mobile Technologies; and has held technical positions at leading companies in these industries, including IBM Research, Akamai, Adobe Omniture, and Qualcomm. His current research is Cloud technologies, with focus on Cloud networking, AIOps, elasticity, and operation efficiency.
\end{minipage}

\begin{minipage}{0.3\linewidth}
    \includegraphics[width=1.5in, clip,keepaspectratio]{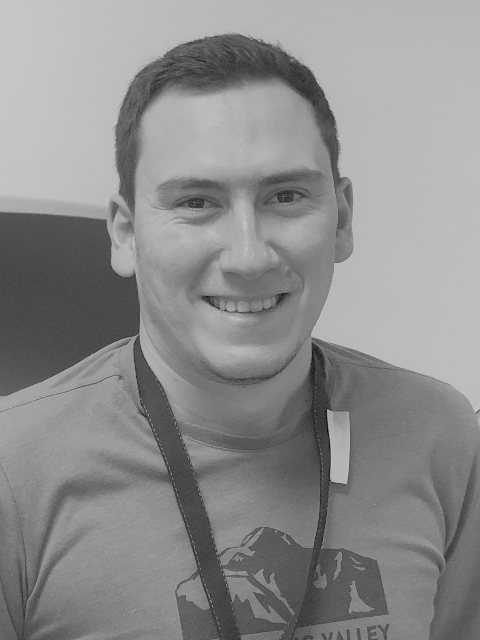}
\end{minipage}\hfil
\begin{minipage}{0.7\linewidth}
    Berkay K\"oksal received his B.Sc. degree in Computer Engineering at Istanbul Technical University and his M.Sc. degree at Eurecom in Internet of Things (IoT) with a focus on data science and smart objects. Currently, he is an ADAS/AD Software Engineer in the Autonomous Vehicle Algorithms (AVA) control team of Renault Software Labs, where his work focuses on production level Vehicle-to-vehicle (V2V) and Vehicle-to-everything (V2X) solutions and system architecture. His research is inspired by data-driven 4G/5G software-defined networks to facilitate and maintain massive IoT network grids in a distributed environment.
\end{minipage}

\begin{minipage}{0.3\linewidth}
    \includegraphics[width=1.5in, clip,keepaspectratio]{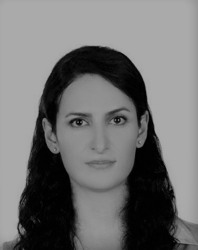}
\end{minipage}\hfil
\begin{minipage}{0.7\linewidth}
    Nasim Ferdosian is a PostDoctoral research fellow in the Communication Systems Department at EURECOM. She was a PostDoctoral researcher at Dublin City University, and an associate researcher with the NEWTON EU H2020 Project in 2017--2018. She received her Ph.D. degree in Computer Science from University Putra Malaysia in 2017. She is currently involved in several European H2020 projects on 5G and network slicing. She is an IEEE member and has served as a Technical Program Committee member for international journals and conferences. Her current research interests are mainly focused on radio resource management, wireless network optimization, artificial intelligence-enabled networking, and machine learning for next-generation wireless network control and management.
\end{minipage}

\end{document}